\documentclass[pdflatex,sn-mathphys-num]{sn-jnl}

\usepackage{graphicx}
\usepackage{amsmath,amssymb,amsthm}
\usepackage{mathtools}
\usepackage{amsfonts}
\usepackage{bm}
\usepackage{xcolor}
\newif\ifshowrevisions
\showrevisionsfalse
\DeclareRobustCommand{\rev}[1]{\ifshowrevisions\begingroup\color{blue}#1\endgroup\else#1\fi}
\newcommand{\revon}{\ifshowrevisions\color{blue}\fi}% use as \begingroup\revon ... \endgroup

\hypersetup{colorlinks=true,linkcolor=black,citecolor=black,urlcolor=black}
\usepackage{enumitem}
\usepackage{booktabs}
\usepackage{textcomp}
\usepackage{comment}
\theoremstyle{thmstyleone}
\newtheorem{theorem}{Theorem}
\newtheorem{proposition}[theorem]{Proposition}
\newtheorem{lemma}[theorem]{Lemma}
\newtheorem{corollary}[theorem]{Corollary}

\theoremstyle{thmstyletwo}
\newtheorem{remark}[theorem]{Remark}
\newtheorem{example}[theorem]{Example}

\theoremstyle{thmstylethree}

\newcommand{\R}{\mathbb{R}}
\newcommand{\1}{\mathbf{1}}
\newcommand{\diag}{\mathrm{diag}}

\newcommand{\norm}[1]{\left\|#1\right\|}

\newcommand{\pos}[1]{\left[#1\right]_+}

\newcommand{\LP}{\mathsf{LP}}

\begin{document}

%\title[Worst-Case Losses in Financial Networks]{Mixed-Integer and Parametric Optimization for Worst-Case Losses in Financial Networks}
\title[Worst-Case Contagion in Financial Networks]{Worst-Case Contagion in Financial Networks: MILP Clearing and Robust Loss Analysis}

\author[1]{\fnm{G.C.} \sur{Calafiore}} %\email{giuseppe.calafiore@polito.it}
\author[1]{\fnm{M.} \sur{Della Rossa}} %\email{matteo.dellarossa@polito.it}
\author*[1]{\fnm{A.} \sur{Proskurnikov}} %\email{anton.proskurnikov@polito.it}

\affil*[1]{\orgdiv{DET}, \orgname{Politecnico di Torino}, \orgaddress{\country{Italy}}}

\abstract{
We study worst-case contagion losses in financial networks with interbank liabilities and common exposures to marketable assets. Existing linear-programming approaches provide exact stress-test values only within an insolvency-resilience margin, where all institutions remain solvent with respect to the external sector. We show that this margin is computed exactly by a finite family of linear programs, one per vertex of the uncertainty ball, whereas a single linear program based on the componentwise worst deterioration of external resources yields only a certified lower bound. We extend this framework beyond that margin in three directions. First, we derive an exact mixed-integer linear programming (MILP) formulation for the maximal clearing vector under arbitrary signed external net positions, providing a clearing oracle for robust stress testing when insolvency can occur. Second, we develop certified bounds on the worst-case loss beyond the margin, including insolvency-pattern enumeration, scenario lower bounds, fixed-pattern dual upper bounds, localized certificates, and a regime-generation procedure with monotone gap certificates. Third, we analyze the worst-case loss curve as a function of the perturbation radius. For $\ell_1$ and $\ell_\infty$ price-perturbation sets, the curve is convex, nondecreasing and piecewise linear up to the margin, with breakpoints generated by switches in worst-case shock directions, LP bases, and bank-level activity patterns. Beyond the margin, it remains nondecreasing and piecewise affine, but convexity may fail, the curve may jump upward, and the worst case need not be attained. The framework provides a certifiable MILP/LP methodology for parametric stress testing and interpretable diagnostics identifying the banks responsible for changes in systemic loss.
}
\keywords{Financial networks; systemic risk; robust stress testing; worst-case loss; mixed-integer linear programming; parametric linear programming; clearing payments.}
\pacs[MSC Classification]{91G45, 90C35, 90C11, 90C17, 90B10}

\maketitle

%%%%%%%%%%%%%%%
\section{Introduction}

Financial networks can transform idiosyncratic balance-sheet shocks into system-wide losses through interbank obligations, overlapping portfolios, and default cascades~\cite{AllenGale2000,FreixasParigiRochet2000,ElliottGolubJackson2014,AcemogluOzdaglarTahbazSalehi2015}. A canonical starting point for quantifying losses from financial contagion is the clearing model of Eisenberg and Noe~\cite{EisenbergNoe2001}, which represents a financial system as a network of nominal liabilities and determines the realized payments under default through a fixed-point clearing equation. The gap between nominal and realized payments then provides a natural measure of system-wide loss induced by external shocks. When the clearing equation admits multiple solutions, the componentwise maximal clearing vector is the canonical selection, as it minimizes total system-wide loss~\cite{EisenbergNoe2001}. This framework has been extended in many directions, including signed external cash flows~\cite{ElsingerLeharSummer2006}, default and bankruptcy costs~\cite{RogersVeraart2013,WeberWeske2017}, fire sales of illiquid assets~\cite{CifuentesFerrucciShin2005,FeinsteinHallaj2023}, multiple maturities~\cite{KusnetsovVeraart2019}, and other features, see surveys~\cite{GlassermanYoung2016,CaccioliBaruccaKobayashi2018}. These models turn systemic-risk questions into well-defined fixed-point and optimization problems, making them particularly suitable for computational analysis of contagion and stress testing~\cite{EisenbergNoe2001,GlassermanYoung2016,CalafioreFracastoroProskurnikov2024}.

When banks may have negative net external positions, clearing and loss evaluation become substantially harder. Some institutions may become insolvent with respect to the external sector, activating discrete insolvency and payment patterns that invalidate the linear-programming characterization used in the solvent-external-sector regime. Mixed-integer linear programming (MILP) provides a natural modeling paradigm for this setting~\cite{AraratMeimanjan2023}: binary variables encode insolvency and payment patterns, while linear constraints describe clearing feasibility conditional on a given pattern.

This computational framework is especially important when shocks are not described by a reliable probability model. In many supervisory stress-testing applications, the analyst may know the size or structure of plausible asset-price deviations, but not their distribution.
In networks with common asset exposures, shocks are not purely bank-specific: a perturbation in the price of one marketable asset can affect all banks holding that asset, thereby generating correlated losses across the system~\cite{HuangLiu2023}.
Computational studies of systemic risk in financial networks typically evaluate contagion under specified, simulated, or empirically calibrated shock scenarios~\cite{Xu2024,TangFan2025,WangZhaoSongShu2026}. A complementary approach is to formulate stress testing as an optimization problem: rather than fixing a shock scenario \emph{ex ante}, the shock is chosen adversarially within a prescribed uncertainty set so as to maximize system-wide loss. Scenario-based analysis asks how a given shock propagates through the network, whereas robust worst-case analysis asks which admissible shock is most damaging.

Robust optimization provides a natural language for this setting: losses are evaluated under the most adverse shock within a prescribed uncertainty set~\cite{BenTalElGhaouiNemirovski2009,BertsimasBrownCaramanis2011}.
A natural uncertainty set is a ball whose radius $\epsilon$ acts as a stress-severity level: fixing a norm on the space of prices and a radius, the worst-case loss is the largest system loss attainable by any asset-price movement of magnitude at most $\epsilon$. Varying $\epsilon$ traces out a loss curve $\eta_{wc}(\epsilon)$ that summarizes how systemic damage grows as the assumed stress intensifies.
 Calafiore et al.~\cite{CalafioreFracastoroProskurnikov2025ANS} introduced this perspective for financial networks with mutual liabilities and common asset holdings, defining a default-resilience margin $\epsilon^\ast$ and a worst-case loss curve $\eta_{wc}(\epsilon)$ under $\ell_1$ and $\ell_\infty$ price perturbations, where $\epsilon$ bounds the magnitude of the asset-price perturbation in the corresponding norm. It was shown in~\cite{CalafioreFracastoroProskurnikov2025ANS} that, up to an insolvency-resilience threshold $\epsilon_{ub}$, worst-case losses can be computed exactly through linear-programming duality.

 From a supervisory standpoint, the value $\epsilon^\ast$ marks the largest shock radius for which the system remains default-free, while \rev{the insolvency-resilience threshold} $\epsilon_{ub}$ \rev{is the largest shock radius for which external-sector insolvency is excluded for every admissible shock, so that} the linear-programming machinery remains valid. The shape of the loss curve between and beyond these thresholds indicates how rapidly systemic losses accumulate as stress intensifies. Kinks of this curve identify stress levels at which the worst-case mechanism changes through a different adverse asset-price direction or a change in the set of defaulting banks.

 Characterizing this curve beyond $\epsilon_{ub}$ is therefore not only a computational problem, but also a prerequisite for interpreting severe-stress scenarios, which are precisely the scenarios supervisors most need to understand. The present paper takes that step by addressing what happens beyond the insolvency-resilience threshold. The restriction $\epsilon \le \epsilon_{ub}$ is not merely technical: \rev{for every $\epsilon>\epsilon_{ub}$ some admissible shock makes a bank externally insolvent, and for such shocks} the linear-programming clearing formulation used in~\cite{CalafioreFracastoroProskurnikov2025ANS} \rev{is infeasible}, although the maximal clearing vector remains well defined. The robust stress-testing problem therefore changes character. It is no longer a purely continuous LP-based computation, but a regime-dependent optimization problem in which banks may switch between positive-payment and zero-payment states. Equivalently, worst-case loss beyond $\epsilon_{ub}$ is governed by a finite family of insolvency regimes and by the lower envelope of their regime-restricted value functions. To analyze this post-insolvency regime, we combine mixed-integer optimization, parametric linear programming, and certified bounding schemes. The resulting framework provides exact clearing formulations, structural descriptions of the worst-case loss curve, and computable lower and upper bounds for systemic-risk stress testing.
 The emphasis is computational: we ask how to evaluate, certify, and interpret worst-case systemic losses when the usual LP formulation ceases to be valid.

\paragraph*{Contributions}

The contributions of this paper are as follows.

First, we provide an exact MILP characterization of the maximal clearing vector for the Eisenberg--Noe model with arbitrary signed external net positions~\cite{ElsingerLeharSummer2006}. Related mixed-integer formulations appear in the systemic-risk literature, especially for Rogers-Veraart clearing mechanisms and set-valued systemic risk measures~\cite{AraratMeimanjan2023}. Our formulation is tailored to the robust common-asset setting and serves as the clearing oracle for all post-insolvency computations.

Second, we represent the loss in the post-insolvency regime in terms of finitely many insolvency and payment patterns. This yields an exact enumeration method for small networks and a practical bounding framework for larger ones, combining scenario-based lower bounds, fixed-pattern upper bounds, regime generation, and localized fixed-pattern certificates.

Third, we study the parametric structure of the worst-case loss curve. On the interval $[0,\epsilon_{ub}]$, where the LP-duality framework is exact, we prove that $\eta_{wc}(\epsilon)$ is a finite upper envelope of affine functions and hence is convex and piecewise linear. When the adverse shock direction is fixed across a breakpoint, the breakpoint-responsible banks are those whose default status changes. Beyond $\epsilon_{ub}$, we show that monotonicity and \rev{a finite piecewise-affine structure} persist for polyhedral uncertainty sets, although convexity is no longer guaranteed once discrete insolvency patterns become active, because the exact loss is a lower envelope over discrete insolvency regimes. \rev{Since the exact loss is only lower semicontinuous in the external positions, the worst-case loss curve may also jump upward, and beyond $\epsilon_{ub}$ the worst case need not be attained.}

The closest predecessor of the present work is~\cite{CalafioreFracastoroProskurnikov2025ANS}, which introduced the robust loss curve for financial networks with common asset holdings and showed how to compute it efficiently up to the insolvency-resilience threshold $\epsilon_{ub}$. The present paper focuses on what lies beyond that LP-solvable regime and on the parametric structure needed to interpret the resulting stress tests. The paper is also related to computational studies of systemic-risk measures and robust financial networks~\cite{AraratMeimanjan2023,HuMitchellTompaidis2024,JarrowProtterQuintos2024}, but differs in its emphasis on common asset-price uncertainty, signed external net positions, mixed-integer clearing beyond insolvency, and explicit radius-dependent loss curves. \rev{We also show that the insolvency-resilience threshold can be computed exactly by solving one LP per vertex of the uncertainty ball (Proposition~\ref{prop:eub-exact}), and that the single LP of~\cite{CalafioreFracastoroProskurnikov2025ANS} provides a certified lower bound on it.}

\paragraph*{Paper organization}
The remainder of the paper is organized as follows. Section~\ref{sec:model} recalls the financial-network model and the definition of maximal clearing vector. Section~\ref{sec:milp} gives the MILP clearing formulation. Section~\ref{sec:post-insolvency} develops post-insolvency robust computation through pattern decomposition and certified bounds. Section~\ref{sec:preub-pwl} studies piecewise linearity and breakpoint diagnostics on the LP-solvable interval. Section~\ref{sec:beyondub} extends the \rev{piecewise-affine} structure beyond the insolvency-resilience radius. Section~\ref{sec:numerics} reports numerical experiments, and Section~\ref{sec:concl} draws conclusions.

%%%%%%%%%%%%%%%%%%%%%%%%%%%%%%%%%%%%%%%%%%%%%%%%%%%%%%%%%%%%%%%%%%%%%
\section{Model, clearing vectors, and loss}\label{sec:model}

This section introduces the financial network model used throughout the paper. The framework combines the Eisenberg--Noe clearing model with signed external net positions~\cite{ElsingerLeharSummer2006} and common asset exposures~\cite{CalafioreFracastoroProskurnikov2025ANS,HuangLiu2023}.

\subsubsection*{Banks and external assets}

Let $V=\{1,\dots,n\}$ be the set of banks and $M=\{1,\dots,m\}$ the set of external assets.
Contractual nominal liabilities are encoded by the matrix $\bar P=(\bar p_{ij})\in\R^{n\times n}$ with $\bar p_{ii}=0$ and $\bar p_{ij}\ge 0$, where $\bar p_{ij}$ is the nominal liability of bank $i$ to bank $j$.
Let $\bar p:=\bar P\1\in\R^n$ be the vector of total nominal interbank liabilities.

Define the row-stochastic matrix of relative liabilities $A=(a_{ij})\in \R^{n\times n}$ by
\begin{equation}
a_{ij}=
\begin{cases}
\bar p_{ij}/\bar p_i,& \bar p_i>0,\\
1,& \bar p_i=0 \text{ and } i=j,\\
0,& \text{otherwise},
\end{cases}
\qquad A\1=\1.
\label{eq:Adef}
\end{equation}
The $i$th row of $A$ encodes how bank $i$'s total nominal obligation is distributed across its creditors. The same proportions apply to actual payments: if bank $i$ makes total interbank payment $p_i \le \bar p_i$, creditor $j$ receives $a_{ij}p_i$. This proportional allocation (pro rata) rule is the standard payment convention in the Eisenberg-Noe framework.
Banks have external net positions $c\in\R^n$ that may be signed. Negative components of $c$ correspond to banks whose external obligations exceed their external assets before interbank clearing takes place. We write these positions as the sum of a sign-unrestricted external cash-flow vector $c^e\in\R^n$ and the marked-to-market value of common asset exposures. Specifically, asset commonality is modeled by an exposure matrix $S\in\R^{n\times m}$ and an asset-price vector $v\in\R^m$, so that
\begin{equation}
c = c^e + Sv.
\label{eq:cdef}
\end{equation}
The entry $S_{ij}$ represents the sensitivity of bank $i$'s balance sheet to the price of asset~$j$ and may be signed. A positive coefficient corresponds to a long exposure, whereas a negative coefficient represents a short exposure or, more generally, a signed factor sensitivity. Thus, a price decline hurts banks with positive exposure and benefits banks with negative exposure, while a price increase has the opposite effect.
This convention is useful in robust stress testing because adverse price directions are selected automatically according to the signs of the relevant exposures.

We consider perturbations $\delta\in\R^m$ around nominal prices $\bar v$:
\begin{equation}
v=\bar v+\delta,\qquad
\bar c := c^e + S\bar v,\qquad
c=\bar c + S\delta,
\label{eq:pert}
\end{equation}
where $\bar c$ is the vector of nominal, unperturbed external net positions and the vector $S\delta$ represents the change in banks' external net positions induced by the asset-price perturbation $\delta$. We assume that $\|\delta\|\leq\epsilon$, where $\epsilon\geq0$ is the perturbation radius and $\|\cdot\|$ is a norm on $\R^m$. The parameter $\epsilon$, called the \emph{perturbation radius}, controls the severity of the stress test: larger values allow more extreme price deviations. Throughout the paper, we work with the $\ell_1$ or the $\ell_\infty$ norm. The $\ell_\infty$ norm bounds each individual asset-price deviation, while the $\ell_1$ norm bounds the total absolute deviation across all assets. Both choices yield polyhedral uncertainty sets, which is the structural property enabling the LP and MILP reformulations developed below.

\subsubsection*{Clearing equation and system loss}

Given $c\in \R^n$, the clearing vector $p\in\R^n$ satisfies the fixed-point equation~\cite{ElsingerLeharSummer2006,AcemogluOzdaglarTahbazSalehi2015,MassaiComoFagnani2022}:
\begin{equation}\label{eq:clearing}
p = \pos{\min(\bar p,\, c + A^\top p)}=\min(\bar p,\pos{c+A^\top p}).
\end{equation}
Here $\min$ is componentwise, and $\pos{\cdot}$ denotes componentwise positive part.
The interpretation of~\eqref{eq:clearing} is as follows.
The quantity $p_i$ is the total interbank payment made by bank $i$. Under the proportionality rule, the bank's available resources consist of its external net position $c_i$ plus incoming interbank receipts $(A^\top p)_i=\sum_j a_{ji}p_j$. Hence bank $i$ pays the smaller of its total nominal obligation $\bar p_i$ and its available resources, subject to limited liability: if available resources are negative, the interbank payment is zero.
This last case corresponds to zero interbank payment caused by negative available resources.
It can occur only when the bank's external net position $c_i$ is negative and incoming interbank receipts are insufficient to compensate;
we call this situation \emph{external-sector insolvency} of bank $i$.

The overall system loss corresponding to the payment vector $p$ is
\begin{equation}
L(p) := \1^\top(\bar p - p).
\label{eq:loss}
\end{equation}
Each component $\bar p_i-p_i$ measures bank $i$'s unpaid interbank liability, and $L(p)$ is the aggregate unpaid interbank liability across the system.

The set of clearing vectors always forms a complete lattice under the componentwise partial order, by the Knaster-Tarski fixed-point theorem~\cite{EisenbergNoe2001,ElsingerLeharSummer2006}. In particular, there exist unique componentwise minimal and maximal clearing vectors. Under certain regularity conditions the clearing vector is unique~\cite{EisenbergNoe2001,CsokaHerings2024,MassaiComoFagnani2022}; otherwise, the minimal and maximal clearing vectors may differ. The \emph{maximal clearing vector}, denoted $p^\ast(c)$, is the economically preferred selection: it represents the largest payments consistent with the clearing equation and therefore minimizes the aggregate unpaid liabilities $L(p)$.
An important computational property is that $p^\ast(c)$ is the unique maximizer of total payments $\1^\top p$ over the set of clearing vectors, or equivalently the unique minimizer of $L(p)$ over that set. This variational characterization allows $p^\ast(c)$ to be computed through LP or MILP formulations, depending on the solvency regime.

Under the uncertainty model introduced in~\eqref{eq:cdef} and~\eqref{eq:pert}, the vector $c$, the maximal clearing vector and the corresponding system loss depend on the perturbation~$\delta$. We are interested in computing the worst-case value of the loss as the perturbation ranges over the uncertainty set $\{\delta:\|\delta\|\leq\epsilon\}$, that is, in the \emph{worst-case loss curve}
\[
\eta_{wc}(\epsilon)
:=
\rev{\sup_{\|\delta\|\leq\epsilon}}
L\bigl(p^\ast(\bar c+S\delta)\bigr),
\]
which records how the largest attainable systemic loss grows as the admissible shock magnitude increases. \rev{We write a supremum because the loss is not continuous in $\delta$ in general: beyond the insolvency-resilience threshold of Section~\ref{subsec:eub}, the supremum need not be attained (Example~\ref{ex:two-bank-cont}). Up to that threshold it is attained (Section~\ref{sec:preub-pwl}).}
We also analyze the associated worst-case perturbations and clearing vectors, which identify the shock directions and institutions responsible for changes in systemic loss.

\subsubsection*{Default-resilience margin}

The analysis below also uses the default-resilience margin introduced in~\cite{CalafioreFracastoroProskurnikov2025ANS}.
Let $\sigma_i^\top$ denote the $i$th row of $S$, denote by $\|\cdot\|_*$ the dual norm associated with $\|\cdot\|$, and let $s$ be the vector having components
$
s_i := \|\sigma_i\|_*$, $i=1,\ldots,n$.
In the uncertainty models used in this paper, $s_i=\|\sigma_i\|_\infty$ for $\ell_1$ perturbations and $s_i=\|\sigma_i\|_1$ for $\ell_\infty$ perturbations, due to the duality between the $\ell_1$ and $\ell_\infty$ norms.

We introduce the vector of nominal residual resources under full payment,
\begin{equation}
\bar r := \bar c + A^\top\bar p - \bar p,
\label{eq:nominal-net-worth}
\end{equation}
so that $\bar r_i$ is the residual resource of bank $i$ after meeting all nominal obligations, assuming full interbank payment. The condition $\bar r \ge 0$ means that every bank can meet its obligations in full at nominal asset prices. Under this nominal no-default condition, the \emph{default-resilience margin}~\cite{CalafioreFracastoroProskurnikov2025ANS} is defined as
\begin{equation}\label{eq:epsilon-star}
\epsilon^\ast := \min_{i:\,\rev{\bar p_i>0,\ }s_i>0} \frac{\bar r_i}{s_i},
\end{equation}
with the usual convention that rows with $s_i=0$ impose no restriction\rev{, and $\epsilon^\ast=+\infty$ if the index set is empty. Banks with $\bar p_i=0$ are excluded because they make no interbank payment and cannot default}.

Indeed, for every perturbation satisfying $\|\delta\|\le \epsilon^\ast$,
\[
(S\delta)_i \ge -\|\sigma_i\|_\ast \|\delta\|
\ge -s_i\epsilon^\ast
\ge -\bar r_i
\]
for all $i$ \rev{with $\bar p_i>0$}, and therefore
\[
\bigl(\bar c+S\delta+A^\top\bar p-\bar p\bigr)_i
=
\bigl(\bar r+S\delta\bigr)_i
\ge 0\qquad\rev{\text{for all $i$ with $\bar p_i>0$.}}
\]
Hence $p^\ast(\bar c+S\delta)=\bar p$ and the worst-case loss is zero. \begingroup\revon Conversely, if $\epsilon>\epsilon^\ast$, let $i$ attain the minimum in~\eqref{eq:epsilon-star} and choose $\delta$ with $\norm{\delta}=\epsilon$ and $\sigma_i^\top\delta=-s_i\epsilon$, which exists because the dual-norm inequality is tight. Then $\bar r_i+\sigma_i^\top\delta<0$, so $\bar p$ is not a clearing vector, $p^\ast(\bar c+S\delta)\neq\bar p$, and the loss is positive.\endgroup{} Thus $\epsilon^\ast$ is the largest perturbation radius \rev{for which} every admissible shock leaves the system default-free; for $\epsilon>\epsilon^\ast$, systemic losses \rev{are positive for some admissible shock}. \rev{This componentwise argument is exact because the condition $p^\ast=\bar p$ fixes the payment vector, so it decouples into one scalar inequality per bank, each attained by its own worst-case shock. The insolvency-resilience threshold of Section~\ref{subsec:eub} involves an adjustable payment vector, and there the same componentwise argument is only sufficient.}

%%%%%%%%%%%%%%%%%%%%%%%%%%%%%%%%%%%%%%%%%%%%%%%%%%%%%%%%%%%%%%%%%%%%%%%%%%%%%%%%%%%%%%%
\section{LP and MILP characterizations of the maximal clearing vector}\label{sec:milp}

As shown in~\cite{CalafioreFracastoroProskurnikov2025ANS}, for a fixed external net-position vector $c=\bar c+S\delta$, the clearing problem admits a linear-programming characterization if there exists a payment vector $p\in[0,\bar p]$ such that $c+A^\top p-p\ge 0$, i.e.,
each bank's available resources, after accounting for interbank receipts, are sufficient to cover its interbank payments. In this case, the maximal clearing vector $p^\ast(c)$ is the unique optimizer of the clearing LP
\begin{equation}\label{eq:feasLP}
\begin{array}{ll}
\displaystyle \min_p & \mathbf 1^\top(\bar p-p)\\[1mm]
\text{s.t.} & 0\le p\le \bar p,\\
& c+A^\top p\ge p.
\end{array}
\end{equation}
In particular, at the maximal clearing vector, every bank's available resources remain nonnegative, even though some banks may still default in the sense that $p_i^\ast(c)<\bar p_i$. This distinguishes the LP-characterizable regime (i.e., when the LP~\eqref{eq:feasLP} is feasible) from the \emph{post-insolvency} regime considered below, where available resources may become negative and the linear-programming characterization breaks down.
Thus LP feasibility excludes the external-sector insolvency, but does not exclude ordinary payment default. \rev{To avoid inessential complications we assume $\bar p>0$, as in all our numerical instances; otherwise the rows of~\eqref{eq:feasLP} corresponding to banks with $\bar p_i=0$ can be dropped, since such banks make no interbank payment and their resources do not affect $p^\ast(c)$.} 
The optimal value in~\eqref{eq:feasLP} is the system loss $L(p^\ast(c))$ at the maximal clearing vector.

\subsection{MILP characterization of the maximal clearing vector}

This subsection provides an exact MILP characterization of $p^\ast(c)$ valid for every $c\in\R^n$, including the post-insolvency regime in which the LP formulation~\eqref{eq:feasLP} is infeasible. Related mixed-integer formulations for computing clearing vectors and systemic-risk measures are available in the literature, notably for Rogers--Veraart-type models with default costs and set-valued risk measures~\cite{AraratMeimanjan2023}. Our formulation is tailored to the framework studied here: it identifies the optimizer with the \emph{maximal} clearing vector, which serves as the clearing oracle for the worst-case loss analysis in the next sections.

Introduce binary variables $z\in\{0,1\}^n$. Given $c\in \R^n$, consider
\begin{equation}
\begin{aligned}
\max_{p\in\mathbb R^n,\ z\in\{0,1\}^n}\quad & \mathbf 1^\top p\\
\text{s.t.}\quad & 0\le p_i\le \bar p_i z_i,\qquad i=1,\dots,n,\\
& p_i\le c_i + (A^\top p)_i + M_i(1-z_i),\qquad i=1,\dots,n,
\end{aligned}
\tag{$\mathrm{MILP}_c$}\label{eq:MILP}
\end{equation}
where $M_i:= \max\{-c_i,0\}$, $i=1,\ldots,n$. For brevity, we refer to the two families of constraints in~\eqref{eq:MILP} as the \emph{box constraints} and the \emph{resource constraints}, respectively.

The value $z_i=1$ selects the branch on which the resource constraint is imposed without relaxation, whereas $z_i=0$ forces zero interbank payment. In particular, $p_i>0$ implies $z_i=1$, but the converse need not hold. 
\begin{theorem}[Exact MILP characterization of the maximal clearing vector]
\label{thm:MILPexact}
For every fixed $c\in\mathbb R^n$, problem \eqref{eq:MILP} is feasible and attains an optimum.
Its optimal payment vector is unique and coincides with the maximal clearing vector $p^\ast(c)$, i.e., the maximal solution of~\eqref{eq:clearing}.
\end{theorem}
\begin{proof}
Notice that $A \geq 0$ by construction and $p \geq 0$ for any feasible pair $(p,z)$. Therefore, $(A^\top p)_i\geq 0$, and hence
\begin{equation}
c_i+(A^\top p)_i+M_i \geq c_i +M_i =  c_i +[-c_i]_+ \geq 0.
\label{eq:Mvalidity}
\end{equation}

\noindent
\textbf{Step 1: every clearing vector induces a feasible MILP point.}
Let $p$ be any clearing vector and define
\[
z_i := \begin{cases}
1,&\text{if }\; c_i+(A^\top p)_i\ge 0,\\
0,&\text{if }\; c_i+(A^\top p)_i<0.
\end{cases}
\]
If $z_i=1$, then the minimum in~\eqref{eq:clearing} is nonnegative, and thus
\[
p_i = \min\{\bar p_i,\, c_i+(A^\top p)_i\},
\]
so the $i$th box and resource constraints hold: $0\le p_i\le \bar p_i$ and $p_i\le  c_i+(A^\top p)_i+M_i(1-z_i)$.

If $z_i=0$, then necessarily $p_i=0$ by \eqref{eq:clearing}, while
\eqref{eq:Mvalidity} gives
\[
0=p_i\le c_i+(A^\top p)_i+M_i=c_i+(A^\top p)_i+M_i(1-z_i).
\]
Thus $(p,z)$ is feasible for~\eqref{eq:MILP}. Since the feasible set is non-empty and compact, the MILP attains an optimum.

\medskip
\noindent\textbf{Step 2: every optimal MILP payment vector is a clearing vector.}
Let $(\tilde p,\tilde z)$ be optimal for~\eqref{eq:MILP}. We are going to prove that $\tilde p$ is a solution to~\eqref{eq:clearing}, that is,
\begin{equation}
\tilde p_i = \theta_i:=\bigl[\min\{\bar p_i,\tilde r_i\}\bigr]_+,
\qquad
\tilde r_i := c_i + (A^\top \tilde p)_i,
\qquad i=1,\dots,n.
\label{eq:patch_target}
\end{equation}

First, $\tilde p_i\leq\theta_i$ follows from the constraints in~\eqref{eq:MILP}. Indeed, if $\tilde z_i=0$, then $\tilde p_i=0\leq\theta_i$ by the box constraint; if $\tilde z_i=1$, then $\tilde p_i\leq\bar p_i$ and $\tilde p_i\leq \tilde r_i$, hence $\tilde p_i\leq\theta_i$.

Suppose, for contradiction, that $\tilde p_i<\theta_i$ for some $i$. Let $(p',z')$ be obtained from $(\tilde p,\tilde z)$ by replacing $\tilde p_i$ with $p_i'=\tilde p_i+\varepsilon$ and $\tilde z_i$ with $z_i'=1$, where $0<\varepsilon<\theta_i-\tilde p_i$. Then the box constraints remain satisfied and $0\leq p'_i=\tilde p_i+\varepsilon<\theta_i\leq\bar p_i=\bar p_i z_i'$.

In particular, $\bar p_i>0$, hence $a_{ii}=0$. The $i$th resource constraint remains feasible, since
\[
p'_i<\theta_i\leq\tilde r_i
=c_i+(A^\top \tilde p)_i
=c_i+(A^\top p')_i+M_i(1-z_i').
\]
For every $k\neq i$, the right-hand side of the $k$th resource constraint is nondecreasing in $p_i$, since increasing $p_i$ increases $(A^\top p)_k$. Hence those constraints also remain satisfied. Therefore $(p',z')$ is feasible and has larger objective value, $\mathbf{1}^\top p'=\mathbf{1}^\top\tilde p+\varepsilon$, contradicting optimality. Thus $\tilde p_i=\theta_i$ for every $i$, which is exactly~\eqref{eq:clearing}. Hence $\tilde p$ is a clearing vector.

\medskip
\noindent\textbf{Step 3: optimality over the MILP equals maximality over clearing vectors.}
Steps 1 and 2 show that every clearing vector is feasible for~\eqref{eq:MILP}, and every MILP optimizer is a clearing vector. Hence~\eqref{eq:MILP} maximizes $\mathbf{1}^\top p$ over the set of clearing vectors.
Since $p^\ast(c)$ componentwise dominates every clearing vector, it is the unique maximizer of $\mathbf{1}^\top p$.
Therefore the optimal payment vector of~\eqref{eq:MILP} is unique and equals $p^\ast(c)$.
\end{proof}

\begin{remark}[Relation to the LP regime]
If the clearing LP~\eqref{eq:feasLP} is feasible, the binary variables in~\eqref{eq:MILP} can be fixed to $z=\1$. The MILP then reduces to the maximization form of~\eqref{eq:feasLP}, since maximizing $\1^\top p$ is equivalent to minimizing $\1^\top(\bar p-p)$.
\label{rem:lp}
\end{remark}

\begin{remark}[Relation to existing MILP formulations]
Formulations structurally related to \eqref{eq:MILP} appear in prior work on systemic-risk computation.
The result of Theorem~\ref{thm:MILPexact} is close in spirit to~\cite[Corollary~4]{AraratMeimanjan2023}. Here we specialize the construction to the present robust Eisenberg--Noe framework with signed external positions, prove uniqueness of the optimizer, and explicitly identify it as the \emph{maximal} clearing vector used throughout Sections~\ref{sec:post-insolvency}--\ref{sec:beyondub}.
\end{remark}

\begin{remark} While the optimal payment vector of~\eqref{eq:MILP} is unique, the associated binary vector need not be: if a bank makes zero payment, $p_i = 0$, then both $z_i = 0$ and $z_i = 1$ can be feasible. None of the subsequent results requires uniqueness of the binary optimizer. 
\end{remark}

\subsection{From fixed-shock clearing to robust stress testing}\label{subsec:eub}

\begingroup\revon
For the $\ell_1$ and $\ell_\infty$ uncertainty sets considered in this paper, the \emph{insolvency-resilience threshold} $\epsilon_{ub}$ is the largest radius $\epsilon$ such that the clearing LP~\eqref{eq:feasLP} with $c=\bar c+S\delta$ is feasible for every perturbation $\norm{\delta}\le\epsilon$; equivalently, by the discussion above, the largest radius for which no admissible shock makes a bank externally insolvent. Let $\mathcal Q$ denote the vertex set of the unit ball $\{\delta:\norm{\delta}\le1\}$, that is, $\mathcal Q=\{\pm e_1,\ldots,\pm e_m\}$ for $\ell_1$ and $\mathcal Q=\{-1,1\}^m$ for $\ell_\infty$.

\begin{proposition}[Exact insolvency-resilience threshold]\label{prop:eub-exact}
Assume $\bar r\ge0$. For $q\in\mathcal Q$ let
\begin{equation}\label{eq:epsq}
\epsilon_q:=\sup_{p,\epsilon}\bigl\{\epsilon\ge0:\ 0\le p\le\bar p,\ \ \bar c+\epsilon Sq+A^\top p\ge p\bigr\}\in[0,+\infty],
\end{equation}
and
\begin{equation}\label{eq:epsub}
\epsilon_{ub}:=\min_{q\in\mathcal Q}\epsilon_q .
\end{equation}
Then, for every $\epsilon\ge0$, the clearing LP~\eqref{eq:feasLP} with $c=\bar c+S\delta$ is feasible for all $\norm{\delta}\le\epsilon$ if and only if $\epsilon\le\epsilon_{ub}$. Moreover, a vertex $q$ can be omitted from~\eqref{eq:epsub} whenever $Sq\ge Sq'$ componentwise for some other $q'\in\mathcal Q$. In particular, if $S\ge0$ it suffices to take $q=-e_j$, $j=1,\ldots,m$, for $\ell_1$ ($m$ LPs) and $q=-\1$ for $\ell_\infty$ (one LP).
\end{proposition}
\begin{proof}
Let $\mathcal F:=\{c\in\R^n:\ \exists p\in[0,\bar p]\ \text{with}\ (I_n-A^\top)p\le c\}$ be the set of external positions for which~\eqref{eq:feasLP} is feasible. It is the projection of a polyhedron, hence a closed convex polyhedron. It is upward closed, since $c\in\mathcal F$ and $c'\ge c$ imply $c'\in\mathcal F$ with the same $p$, and $\bar c\in\mathcal F$, since $p=\bar p$ is feasible when $\bar r\ge0$. For fixed $q$, the set $\{\epsilon\ge0:\ \bar c+\epsilon Sq\in\mathcal F\}$ is therefore a closed interval containing $0$, namely $[0,\epsilon_q]$ (or $[0,\infty)$ if $\epsilon_q=+\infty$).
If $\epsilon\le\epsilon_{ub}$, then $\bar c+\epsilon Sq\in\mathcal F$ for every $q\in\mathcal Q$. Every $\delta$ with $\norm{\delta}\le\epsilon$ is a convex combination $\delta=\sum_{q}\theta_q\,\epsilon q$ of the scaled vertices, so $\bar c+S\delta=\sum_q\theta_q(\bar c+\epsilon Sq)\in\mathcal F$ by convexity. Conversely, if $\epsilon>\epsilon_{ub}$ and $q^\ast$ attains the minimum in~\eqref{eq:epsub}, the admissible shock $\delta=\epsilon q^\ast$ gives $\bar c+S\delta\notin\mathcal F$.
Finally, if $Sq\ge Sq'$, upward closedness gives $\bar c+\epsilon Sq\in\mathcal F$ whenever $\bar c+\epsilon Sq'\in\mathcal F$, hence $\epsilon_q\ge\epsilon_{q'}$ and $q$ does not affect the minimum.
\end{proof}

Each $\epsilon_q$ is the optimal value of an LP of the same size as~\eqref{eq:feasLP}; equivalently, $\epsilon_{ub}$ is the optimal value of a single LP with one payment vector $p^{(q)}$ per vertex. For $\ell_1$ uncertainty this requires $2m$ LPs ($m$ for long-only exposures). For $\ell_\infty$ uncertainty with general signed exposures $|\mathcal Q|=2^m$, but a single vertex suffices when $S$ is sign-consistent, i.e., $SD\ge0$ for some $D=\diag(d)$ with $d\in\{-1,1\}^m$: then $q=-d$ dominates all other vertices.

\paragraph*{A certified lower bound}
The single LP of~\cite[Theorem~3]{CalafioreFracastoroProskurnikov2025ANS},
\begin{equation}\label{eq:epsub-lb}
\begin{array}{ll}
\displaystyle \underline\epsilon_{ub}= \max_{p,\epsilon} & \epsilon \\[1mm]
\text{s.t.} & 0\le p\le \bar p,\\
& \bar c - \epsilon s +A^\top p\ge p,\\
& \epsilon\ge0,
\end{array}
\end{equation}
where $s_i=\|\sigma_i\|_\ast$ as defined in Section~\ref{sec:model}, replaces all admissible positions by the componentwise worst deterioration: $\bar c+S\delta\ge\bar c-\epsilon s$ for every $\norm{\delta}\le\epsilon$, by the dual-norm inequality. Since $\mathcal F$ is upward closed, feasibility at $\bar c-\epsilon s$ implies feasibility for every admissible shock, so~\eqref{eq:epsub-lb} is a certified lower bound,
\[
\underline\epsilon_{ub}\le\epsilon_{ub}.
\]
The inequality is strict in general. Each row bound $(S\delta)_i\ge-s_i\epsilon$ is attained, but by a shock that depends on~$i$, so the point $\bar c-\epsilon s$ is usually not itself admissible. Equality holds whenever one vertex attains all row-wise worst cases simultaneously, i.e., $Sq=-s$ for some $q\in\mathcal Q$. This is the case for $\ell_\infty$ uncertainty with sign-consistent $S$ (and in particular for $S\ge0$), but in general not for $\ell_1$ uncertainty with $m\ge2$. On three of the five instances of Section~\ref{sec:numerics}, $\epsilon_{ub}$ exceeds $\underline\epsilon_{ub}$ by $5.5\%$, $10\%$ and $68\%$ (Table~\ref{tab:instances}). In contrast, the componentwise argument is exact for the default-resilience margin $\epsilon^\ast$, where the payment vector is fixed at $\bar p$.

\begin{example}[Two-bank cycle]\label{ex:two-bank}
Let two banks owe one unit to each other, so that $\bar p=(1,1)^\top$ and $A=\left[\begin{smallmatrix}0&1\\1&0\end{smallmatrix}\right]$, and let $\bar c=(1,\tfrac12)^\top$, $S=\diag(2,1)$, with $\ell_1$ uncertainty. Then $\bar r=\bar c$, $s=(2,1)^\top$ and $\epsilon^\ast=\tfrac12$. Writing $d=p_1-p_2\in[-1,1]$, the LP~\eqref{eq:feasLP} is feasible if and only if $c_1+c_2\ge0$, $c_1\ge-1$ and $c_2\ge-1$. The vertex $q=-e_1$ gives $c=(1-2\epsilon,\tfrac12)^\top$ and $\epsilon_{-e_1}=\tfrac34$; the vertex $q=-e_2$ gives $c=(1,\tfrac12-\epsilon)^\top$ and $\epsilon_{-e_2}=\tfrac32$. Hence $\epsilon_{ub}=\tfrac34$. The lower bound~\eqref{eq:epsub-lb} tests $\bar c-\epsilon s=(1-2\epsilon,\tfrac12-\epsilon)^\top$ and gives $\underline\epsilon_{ub}=\tfrac12$: for $\epsilon\in(\tfrac12,\tfrac34]$ the clearing LP is feasible for every admissible shock, although~\eqref{eq:epsub-lb} cannot certify it. We return to this example in Section~\ref{sec:beyondub}.
\end{example}
\endgroup

The fixed-shock clearing problem should be distinguished from the robust stress-testing problem, in which the perturbation $\delta$ is itself chosen adversarially within the uncertainty set. For a given $\epsilon\le\epsilon_{ub}$, the worst-case loss $\eta_{wc}$ can be computed by the LP-duality approach of~\cite{CalafioreFracastoroProskurnikov2025ANS}. Section~\ref{sec:preub-pwl} gives a full parametric characterization of this loss curve, proving piecewise linearity and identifying the mechanisms that generate its breakpoints. \rev{For every $\epsilon>\epsilon_{ub}$, by Proposition~\ref{prop:eub-exact} there are admissible perturbations for which}~\eqref{eq:feasLP} is infeasible. At the corresponding exact clearing vector, some banks may have negative total available resources and therefore make zero interbank payments. The fixed-point clearing equation remains well defined, but its LP characterization no longer applies. This is precisely the regime for which the MILP formulation developed above is needed.

%%%%%%%%%%%%%%%%%%%%%%%%%%%%%%%%%%%%%%%%%%%%%%%%%%%%%%%%%%%%%%%%%%%%%%%%%%%%%%%%%%%%%%%%%%%%%%%%%%%%%%%%%%%%%%%
\section{Worst-case losses beyond \texorpdfstring{$\epsilon_{ub}$}{epsilon ub}}\label{sec:post-insolvency}

This section develops certified methods for computing the post-insolvency worst-case loss curve defined in Section~\ref{sec:model}:
\begin{equation}
\begin{gathered}
\eta_{wc}(\epsilon)=\rev{\sup_{\norm{\delta}\le \epsilon}}\eta^\ast\!(c(\delta)), \;\text{ with }\;
c(\delta):=\bar c+S\delta,\\
\eta^\ast(c):=L\bigl(p^\ast(c)\bigr)=\1^\top(\bar p-p^\ast(c)).
\end{gathered}
\label{eq:wcdef}
\end{equation}
For a fixed perturbation $\delta$, evaluating the objective in~\eqref{eq:wcdef} requires computing $p^\ast(\bar c+S\delta)$.
For $\epsilon\le \epsilon_{ub}$ this inner evaluation can still be handled by the LP-duality machinery developed in \cite{CalafioreFracastoroProskurnikov2025ANS}. For $\epsilon>\epsilon_{ub}$, \rev{some} admissible perturbations \rev{lead} to external-sector insolvency \rev{(Proposition~\ref{prop:eub-exact})}, and exact inner clearing must in general use the MILP oracle of Theorem~\ref{thm:MILPexact}. In this case, robust stress testing has a bilevel structure: an outer maximization over admissible shocks is coupled with an inner exact clearing computation that returns the clearing vector $p^\ast(c(\delta))$ and the corresponding loss $\eta^\ast(c(\delta))$.

To this end, we express the maximal clearing value $\mathbf 1^\top p^\ast(c)$ as the maximum over finitely many regime-restricted LP values, or equivalently express the exact loss $\eta^\ast(c)$ as the minimum over finitely many regime-restricted loss values. Each regime is indexed by a binary insolvency/payment pattern. The next subsection makes this decomposition precise by defining, for each pattern $z\in\{0,1\}^n$, a regime-restricted feasible set, payment value, and loss value. After deriving this exact but exponential representation, we develop certified lower and upper bounds and combine them in a regime-generation scheme with monotone gap certificates.

\subsection{Pattern decomposition of the exact loss}

Let $z\in\{0,1\}^n$ be a binary insolvency/payment pattern, where $z_i=1$ indicates that bank $i$ is allowed to make positive interbank payments and $z_i=0$ forces $p_i=0$; we set $Z:=\diag(z)$.

For fixed $z$ and $c$, consider the regime-restricted polyhedron
\begin{equation}
\mathcal P_z(c)
:=
\Bigl\{p\in\R^n:
0\le p\le Z\bar p,\qquad
Zp\le Z\bigl(c+A^\top p\bigr)
\Bigr\}.
\label{eq:Pz-clean}
\end{equation}
Every $p\in\mathcal P_z(c)$ gives a feasible pair $(p,z)$ for~\eqref{eq:MILP}.
Indeed, the box constraints are satisfied by definition of $\mathcal P_z(c)$: $0\le p_i\le \bar p_i z_i$ for all $i$.
If $z_i=1$, then $p_i\le c_i+(A^\top p)_i$.
If $z_i=0$, then $p_i=0$ and thus, in view of $A^\top p\geq 0$, one has
\[
0=p_i\le c_i+(A^\top p)_i+(-c_i)_+=c_i+(A^\top p)_i+M_i.
\]
Therefore, the $i$th resource constraint of the MILP is also satisfied.

For fixed $z$, the value
\begin{equation}
v_z(c):=\max\{\1^\top p: p\in\mathcal P_z(c)\}
\label{eq:vzdef}
\end{equation}
is the maximal total payment compatible with pattern $z$, with the convention $v_z(c)=-\infty$ when $\mathcal P_z(c)=\varnothing$.
We also define the associated regime-restricted loss
\begin{equation}
\eta_z(c):=\1^\top\bar p-v_z(c),
\label{eq:etazdef}
\end{equation}
with the convention $\eta_z(c)=+\infty$ when $\mathcal P_z(c)=\varnothing$.

\begin{proposition}[Pattern decomposition of the exact clearing value and loss]
For all $c\in\R^n$,
\begin{equation}
\1^\top p^\ast(c)=\max_{z\in\{0,1\}^n} v_z(c),
\qquad
\eta^\ast(c)=\min_{z\in\{0,1\}^n}\eta_z(c).
\label{eq:patterndecomp}
\end{equation}
\label{prop:patterndecomp}
\end{proposition}
\begin{proof}
Let $p^\ast(c)$ be the maximal clearing vector and define the binary vector $z^\ast$ by
$z_i^\ast=1$ whenever $p_i^\ast(c)>0$ and $z_i^\ast=0$ whenever $p_i^\ast(c)=0$.
Then $0\le p^\ast(c)\le Z^\ast\bar p$.
Moreover, if $z_i^\ast=1$, then $p_i^\ast(c)>0$, and the clearing equation implies
\[
p_i^\ast(c)=\min\{\bar p_i,\ c_i+(A^\top p^\ast(c))_i\}\le c_i+(A^\top p^\ast(c))_i.
\]
Hence $Z^\ast p^\ast(c)\le Z^\ast(c+A^\top p^\ast(c))$, so $p^\ast(c)\in\mathcal P_{z^\ast}(c)$.
Therefore
\[
\max_{z\in\{0,1\}^n} v_z(c)\ge v_{z^\ast}(c)\ge \1^\top p^\ast(c).
\]

Conversely, fix any $z\in\{0,1\}^n$ and any $p\in\mathcal P_z(c)$.
Since $(p,z)$ is feasible for the exact MILP of Theorem~\ref{thm:MILPexact}, one has
$\1^\top p\le \1^\top p^\ast(c)$. Taking the maximum over $p\in\mathcal P_z(c)$ and then over $z$ yields
the inverse inequality:
\[
\max_{z\in\{0,1\}^n} v_z(c)\le \1^\top p^\ast(c).
\]
Combining the two inequalities proves the first identity in \eqref{eq:patterndecomp};
the second follows by subtracting both sides from $\1^\top\bar p$.
\end{proof}

Proposition~\ref{prop:patterndecomp} immediately yields the exact post-insolvency robust reformulation
\begin{equation}
\eta_{wc}(\epsilon)
=
\rev{\sup_{\norm{\delta}\le\epsilon}}\ \min_{z\in\{0,1\}^n}\eta_z(\bar c+S\delta).
\label{eq:wc-pattern}
\end{equation}
This identity makes the role of the pattern decomposition explicit: it converts the outer robust problem into the maximization of a lower envelope of finitely many pattern-restricted LP value functions.

Proposition~\ref{prop:patterndecomp} also gives an exact but exponential procedure for a \emph{fixed} vector $c$: when $n$ is small, one may evaluate all $2^n$ patterns, solve the LP~\eqref{eq:vzdef} for each feasible pattern, and then select the pattern with maximal total payment, equivalently minimal loss. For robust worst-case analysis over $\norm{\delta}\le\epsilon$, a direct enumeration of all patterns inside the outer maximization is practical only for very small networks.
This motivates the certified approximation schemes developed next.

\subsection{Scalable certified bounds}\label{subsec:certified-bounds}

We next develop a certified bounding framework for the worst-case loss value $\eta_{wc}(\epsilon)$. Exact evaluations at finitely many admissible scenarios provide deterministic lower bounds (Section~\ref{subsec:scenario-lower}), while fixed insolvency/payment patterns provide robust upper bounds through regime-restricted LP duality (Section~\ref{subsec:fixed-pattern-upper}).  These certificates are combined in the regime-generation procedure of Section~\ref{subsec:regime-localized}, which yields monotonically tightening gaps.

\subsubsection{Scenario-based lower bounds}\label{subsec:scenario-lower}

Let $\Delta=\{\delta^{(1)},\dots,\delta^{(K)}\}$ be any finite subset of the uncertainty set
\[
\mathcal U_\epsilon:=\{\delta\in\R^m:\ \norm{\delta}\le\epsilon\}.
\]
We write $|\Delta|=K$ for its cardinality and call $\Delta$ a finite scenario set. For each sampled scenario $\delta^{(k)}\in \Delta$, 
compute the exact loss using the MILP of Theorem~\ref{thm:MILPexact}:
\[
\eta^{(k)}:=\eta^\ast(c^{(k)}),\quad c^{(k)}:=\bar c+S\delta^{(k)}.
\]
For an arbitrary scenario set, the value
\begin{equation}
LB_\Delta(\epsilon):=\max_{k\in \{1,\dots,|\Delta|\}}\eta^{(k)}
\label{eq:sampleLB}
\end{equation}
serves as a natural lower bound for the worst-case loss value $\eta_{wc}(\epsilon)$.
\begin{proposition}[Deterministic scenario lower bound]
For every finite scenario set $\Delta\subseteq\mathcal U_\epsilon$,
\begin{equation}
LB_\Delta(\epsilon)\le \eta_{wc}(\epsilon).
\label{eq:sample-lb-valid}
\end{equation}
Moreover, for any finite scenario set $\Delta_1\subseteq \mathcal U_\epsilon$ such that  $\Delta_1\supseteq \Delta$, we have
\begin{equation}
LB_{\Delta_1}(\epsilon)\ge LB_\Delta(\epsilon).
\label{eq:sample-lb-mon}
\end{equation}
\label{prop:samplelb}
\end{proposition}

\begin{proof}
Given  $\Delta=\{\delta^{(1)},\dots,\delta^{(K)}\}$, since every $\delta^{(k)}$ belongs to $\mathcal U_\epsilon$, we have
\[
\eta^{(k)}=\eta^\ast(\bar c+S\delta^{(k)})
\le
\rev{\sup_{\norm{\delta}\le\epsilon}}\eta^\ast(\bar c+S\delta)=\eta_{wc}(\epsilon)
\]
for each $k$.
Taking the maximum over $k$ proves \eqref{eq:sample-lb-valid}.
Monotonicity \eqref{eq:sample-lb-mon} is immediate because enlarging the scenario set can only enlarge the maximum.
\end{proof}

The scenario set may be random, quasi-random, or fully adversarial.
Perturbations returned by deterministic stress directions, random or quasi-random sampling, fixed-pattern searches, or the regime-generation scheme below (Section~\ref{subsec:regime-localized}) can be inserted into $\Delta$ without affecting validity.
The guarantee in Proposition~\ref{prop:samplelb} is deterministic and does not require any probabilistic assumptions.

\subsubsection{Fixed-pattern certified upper bounds}\label{subsec:fixed-pattern-upper}

For upper bounds, the key observation is that for fixed $z\in\{0,1\}^n$, the restricted value $v_z(c)$ is an LP value function and can therefore be robustified analytically. We first derive a dual representation.

\begin{lemma}[Dual of the regime-restricted LP]
Fix $z\in\{0,1\}^n$ and let $Z=\diag(z)$.
Whenever $\mathcal P_z(c)\neq\varnothing$, the value $v_z(c)$ in \eqref{eq:vzdef} is also given by
\begin{equation}
v_z(c)
=
\min_{\beta,\lambda\ge0}
\Bigl\{(Z\bar p)^\top\beta+(Zc)^\top\lambda\; :\;
\beta+(I_n-A)Z\lambda\ge\1\Bigr\}.
\label{eq:vz-dual}
\end{equation}
Equivalently,
\begin{equation}
\eta_z(c)
=
\max_{\beta,\lambda\ge0}
\Bigl\{
\1^\top\bar p-(Z\bar p)^\top\beta-(Zc)^\top\lambda\; :\;
\beta+(I_n-A)Z\lambda\ge\1
\Bigr\}.
\label{eq:etaz-dual}
\end{equation}
\label{lem:vzdual}
\end{lemma}

\begin{proof}
The primal problem \eqref{eq:vzdef} can be written as
\[
\max_p\ \1^\top p
\quad\text{s.t.}\quad
p\le Z\bar p,
\qquad
Z(I-A^\top)p\le Zc,
\qquad
p\ge0.
\]
Its feasible set is bounded by $0\le p\le Z\bar p$.
Associate dual multipliers $\beta\ge0$ with $p\le Z\bar p$ and $\lambda\ge0$ with
$Z(I_n-A^\top)p\le Zc$.
The dual is therefore
\[
\min_{\beta,\lambda\ge0}
(Z\bar p)^\top\beta+(Zc)^\top\lambda
\quad\text{s.t.}\quad
\beta+\bigl[Z(I_n-A^\top)\bigr]^\top\lambda\ge\1.
\]
Since $Z$ is diagonal, $\bigl[Z(I_n-A^\top)\bigr]^\top=(I_n-A)Z$, leading to~\eqref{eq:vz-dual}.
Strong duality holds because the primal is feasible and bounded.
Subtracting the optimal value from $\1^\top\bar p$ yields \eqref{eq:etaz-dual}.
\end{proof}

For a fixed pattern $z$, define its robust worst-case bound
\begin{equation}
UB_z(\epsilon):=
\max_{\norm{\delta}\le\epsilon}\eta_z(\bar c+S\delta),
\label{eq:UBzdef}
\end{equation}
with the convention $UB_z(\epsilon)=+\infty$ if $\mathcal P_z(\bar c+S\delta)=\varnothing$ for some admissible $\delta$.

\begin{proposition}[Fixed-pattern upper bounds]
For every $z\in\{0,1\}^n$,
\begin{equation}
\eta_{wc}(\epsilon)\le UB_z(\epsilon).
\label{eq:ubz-valid}
\end{equation}
Hence, for every finite family $\mathcal Z\subseteq\{0,1\}^n$,
\begin{equation}
\eta_{wc}(\epsilon)
\le
UB_{\mathcal Z}(\epsilon)
:=
\min_{z\in\mathcal Z} UB_z(\epsilon).
\label{eq:family-ub-valid}
\end{equation}
Moreover, whenever $\mathcal P_z(\bar c+S\delta)\neq\varnothing$ for all $\delta\in\mathcal U_\epsilon$, the quantity $UB_z(\epsilon)$ admits the exact dual representation
\begin{equation}
UB_z(\epsilon)
=
\max_{\beta,\lambda\ge0}
\Bigl\{
\1^\top\bar p-(Z\bar p)^\top\beta-(Z\bar c)^\top\lambda
+\epsilon\,\norm{S^\top Z\lambda}_\ast\; :\;
\beta+(I_n-A)Z\lambda\ge\1
\Bigr\}.
\label{eq:UBz-dual}
\end{equation}
\label{prop:patternub}
\end{proposition}

\begin{proof}
From Proposition~\ref{prop:patterndecomp}, for every $c$ we have
\[
\eta^\ast(c)=\min_{\tilde z\in\{0,1\}^n}\eta_{\tilde z}(c)\le \eta_z(c).
\]
Evaluating this inequality at $c=\bar c+S\delta$ and maximizing over $\norm{\delta}\le\epsilon$ gives \eqref{eq:ubz-valid}.
Taking the minimum over $z\in\mathcal Z$ yields \eqref{eq:family-ub-valid}.

Assume now that $\mathcal P_z(\bar c+S\delta)$ is nonempty for all $\delta\in\mathcal U_\epsilon$.
By Lemma~\ref{lem:vzdual},
\[
\eta_z(\bar c+S\delta)
=
\max_{\beta,\lambda\ge0}
\Bigl\{
\1^\top\bar p-(Z\bar p)^\top\beta-(Z\bar c)^\top\lambda-\delta^\top S^\top Z\lambda:
\beta+(I_n-A)Z\lambda\ge\1
\Bigr\}.
\]
Maximizing over $\norm{\delta}\le\epsilon$ and using
\[
\max_{\norm{\delta}\le\epsilon}(-\delta^\top a)=\epsilon\norm{a}_\ast
\qquad\forall a\in\R^m
\]
yields \eqref{eq:UBz-dual}.
\end{proof}

The all-zero pattern $z=0$ is always feasible with $\mathcal{P}_0(c)=\{0\}$ and gives the trivial but finite upper bound
\[
UB_{0}(\epsilon)=\1^\top\bar p.
\]
Hence the family bound \eqref{eq:family-ub-valid} is always well defined as soon as the zero pattern is included.
Patterns for which $UB_z(\epsilon)=+\infty$ are harmless: they can simply be discarded because they never improve the minimum in~\eqref{eq:family-ub-valid}.

\subsubsection{Tractable specializations for $\ell_1$ and $\ell_\infty$ uncertainty}

The finite LP reductions of \eqref{eq:UBz-dual} follow from the
polyhedral support-function representations of the dual norms associated
with the two uncertainty sets used in this paper. The issue is not whether
$S$ is signed per se, but whether the sign of $S^\top Z\lambda$ is known
a priori. When $S$ has arbitrary signs, this vector is sign-indeterminate,
and both positive and negative shock directions must be retained.

Consider the sets of extreme points for the $\ell_{\infty}$ and $\ell_1$ balls, respectively:
\begin{equation}\label{eq:unit-balls}
\mathcal Q_\infty:=\{-1,1\}^m,
\qquad
\mathcal Q_1:=\{\pm e_1,\ldots,\pm e_m\}.
\end{equation}
Then the following reductions hold:
\begin{itemize}
\item\ If the uncertainty norm is $\ell_\infty$, its dual norm is $\ell_1$, and
\[
\norm{S^\top Z\lambda}_1
=
\max_{q\in\mathcal Q_\infty} q^\top S^\top Z\lambda .
\]
Therefore \eqref{eq:UBz-dual} is the maximum of $2^m$ linear-programming
values, one for each signed joint asset-movement direction $q\in
\mathcal Q_\infty$.

\item\ If the uncertainty norm is $\ell_1$, its dual norm is $\ell_\infty$, and
\[
\norm{S^\top Z\lambda}_\infty
=
\max_{q\in\mathcal Q_1} q^\top S^\top Z\lambda .
\]
Therefore \eqref{eq:UBz-dual} is the maximum of $2m$ linear-programming
values, one for each signed single-asset direction.
\end{itemize}

In the long-only case $S\ge0$, since $Z\lambda\ge0$ for every
$\lambda\ge0$, we have $S^\top Z\lambda\ge0$. Hence the above finite
families simplify: in the $\ell_\infty$ case only the all-positive
direction $q=\mathbf 1$ is needed, while in the $\ell_1$ case only the
$m$ positive single-asset directions $q=e_j$, $j=1,\ldots,m$, are needed.

\subsection{Regime generation and localized certificates}\label{subsec:regime-localized}

\begingroup
The certificates developed in Section~\ref{subsec:certified-bounds} can be tightened adaptively. We combine scenario lower bounds and fixed-pattern upper bounds in a regime-generation loop with monotone gap certificates (Section~\ref{subsec:gap-certificate}), and then localize the fixed-pattern upper bounds over a polyhedral partition of the uncertainty set (Section~\ref{subsec:localized-pattern-ub}).

%We did not use the continuous relaxation of the exact MILP as a robust optimization oracle. For a fixed vector $c$ the binary relaxation is a linear program, but after substituting $c=\bar c+S\delta$ into the robust search, the terms $(-c_i)_+(1-z_i)$ become nonlinear in $(\delta,z)$. Its robust maximization is therefore not an LP and was not used in the numerical experiments. 
In the certified scheme below, candidate shocks are generated by deterministic or sampled scenarios, fixed-pattern searches, and local cell-wise searches; every reported lower bound is certified only after exact clearing is recomputed at the generated shock.
\endgroup

\subsubsection{Regime generation and gap certificates}
\label{subsec:gap-certificate}

The lower and upper bounds above suggest a natural refinement loop.
At each step $k$, maintain two finite sets:
\begin{itemize}[leftmargin=1.8em]
\item a scenario set $\Delta_k\subseteq\mathcal U_\epsilon$;
\item a pattern set $\mathcal Z_k\subseteq\{0,1\}^n$, always containing $0$.
\end{itemize}
Define the associated bounds
\begin{equation}
LB_k(\epsilon):=LB_{\Delta_k}(\epsilon)=
\max_{\delta\in\Delta_k}\eta^\ast(\bar c+S\delta),
\qquad
UB_k(\epsilon):=UB_{\mathcal Z_k}(\epsilon)=
\min_{z\in\mathcal Z_k} UB_z(\epsilon).
\label{eq:LBUBk}
\end{equation}
A basic regime-generation step is:
\begin{enumerate}[leftmargin=1.9em,label=\arabic*.]
\item choose a pattern
\[
z^{(k)}\in\arg\min_{z\in\mathcal Z_k} UB_z(\epsilon);
\]
\item solve the robust fixed-pattern problem for $z^{(k)}$ and extract one optimizer $\delta^{(k)}\in\mathcal U_\epsilon$;
\item solve an exact clearing oracle at $c^{(k)}=\bar c+S\delta^{(k)}$ (for instance, the MILP of Theorem~\ref{thm:MILPexact}), obtaining the exact loss
\[
\eta^{(k)}:=\eta^\ast(c^{(k)})
\]
and selecting an activity pattern $\hat z^{(k)}$ compatible with the exact clearing vector;
\item update
\[
\Delta_{k+1}=\Delta_k\cup\{\delta^{(k)}\},
\qquad
\mathcal Z_{k+1}=\mathcal Z_k\cup\{\hat z^{(k)}\}.
\]
\end{enumerate}

\begin{proposition}[Monotone tightening and gap certificates]
For every iteration $k$,
\begin{equation}
LB_k(\epsilon)\le \eta_{wc}(\epsilon)\le UB_k(\epsilon).
\label{eq:gap-cert}
\end{equation}
Moreover,
\begin{equation}
LB_{k+1}(\epsilon)\ge LB_k(\epsilon),
\qquad
UB_{k+1}(\epsilon)\le UB_k(\epsilon).
\label{eq:monotone-tightening}
\end{equation}
Finally, if at some iteration $k$ the selected scenario satisfies
\begin{equation}
\eta^\ast(\bar c+S\delta^{(k)})=UB_{z^{(k)}}(\epsilon),
\label{eq:exact-hit}
\end{equation}
then
\begin{equation}
LB_{k+1}(\epsilon)=\eta_{wc}(\epsilon)=UB_{k+1}(\epsilon)=UB_k(\epsilon),
\label{eq:finite-hit}
\end{equation}
so the algorithm terminates with an exact certificate.
\label{prop:regimegen}
\end{proposition}
\begin{proof}
The bound \eqref{eq:gap-cert} follows directly from Propositions~\ref{prop:samplelb} and \ref{prop:patternub} applied to the sets $\Delta_k$ and $\mathcal Z_k$.
Monotonicity in \eqref{eq:monotone-tightening} follows from the inclusions
$\Delta_k\subseteq\Delta_{k+1}$ and $\mathcal Z_k\subseteq\mathcal Z_{k+1}$.

Assume now that \eqref{eq:exact-hit} holds.
Then, by construction of $z^{(k)}$,
\[
UB_k(\epsilon)=UB_{z^{(k)}}(\epsilon)=\eta^\ast(\bar c+S\delta^{(k)}).
\]
Since $\delta^{(k)}\in\Delta_{k+1}$, we also have
\[
LB_{k+1}(\epsilon)
\ge
\eta^\ast(\bar c+S\delta^{(k)})
=
UB_k(\epsilon).
\]
Combining this with~\eqref{eq:gap-cert} and~\eqref{eq:monotone-tightening} gives
\[
UB_k(\epsilon)
\leq
LB_{k+1}(\epsilon)
\leq
\eta_{wc}(\epsilon)\leq UB_{k+1}(\epsilon)\leq
UB_k(\epsilon),
\]
which proves \eqref{eq:finite-hit}.
\end{proof}

\begingroup\revon
\begin{remark}[Non-attainment and stopping rule]\label{rem:nonattain}
Equality~\eqref{eq:finite-hit} implies that the supremum defining $\eta_{wc}(\epsilon)$ is attained at $\delta^{(k)}$. For $\epsilon\le\epsilon_{ub}$ this is always the case, because $\eta^\ast$ is continuous on the LP-feasible set (Proposition~\ref{prop:post-c-pwl}). Beyond $\epsilon_{ub}$, however, the supremum need not be attained (Example~\ref{ex:two-bank-cont}). In that case $LB_k(\epsilon)<\eta_{wc}(\epsilon)$ for every $k$, since each $LB_k(\epsilon)$ is the loss of one admissible scenario, so the gap $UB_k(\epsilon)-LB_k(\epsilon)$ never vanishes and exact termination is impossible. In practice, the procedure is stopped when the relative gap falls below a prescribed tolerance, which certifies $\eta_{wc}(\epsilon)$ to within that tolerance; an exact certificate~\eqref{eq:finite-hit} certifies in addition that the worst case is attained.
\end{remark}
\endgroup

Proposition~\ref{prop:regimegen} yields a practical stopping rule through the gap
$UB_k(\epsilon)-LB_k(\epsilon)$.

\subsubsection{Localized fixed-pattern certificates}\label{subsec:localized-pattern-ub}

The fixed-pattern certificate in Proposition~\ref{prop:patternub} is global in the sense that a pattern
$z$ must be feasible for every perturbation in the whole uncertainty set $\mathcal U_\epsilon$ in order to
produce a finite value $UB_z(\epsilon)$.  This requirement can be conservative when $\epsilon$ is large:
a payment regime that is invalid in one part of $\mathcal U_\epsilon$ may still give a sharp certificate in
another part.  A direct way to reduce this conservatism is to localize the fixed-pattern certificate.
Let
\begin{equation}\label{eq:finiteCover}
        \mathcal U_\epsilon=\bigcup_{\ell=1}^C \mathcal U_\epsilon^\ell
\end{equation}
be a finite cover of the uncertainty set by compact polytopes, referred to as \emph{cells}.  For a pattern $z$ and a cell
$\mathcal U_\epsilon^\ell$, define
\[
        UB_{z,\ell}(\epsilon)
        :=
        \max_{\delta\in\mathcal U_\epsilon^\ell}
        \eta_z(\bar c+S\delta),
\]
with the convention $UB_{z,\ell}(\epsilon)=+\infty$ if
$\mathcal P_z(\bar c+S\delta)$ is empty for some
$\delta\in\mathcal U_\epsilon^\ell$.

The fixed-pattern upper bound from Proposition~\ref{prop:patternub} admits a localized counterpart.
\begin{proposition}[Localized fixed-pattern upper bound]\label{prop:patternub-local}
Given a finite cover of the uncertainty set $\mathcal U_\epsilon$ as in~\eqref{eq:finiteCover}, fix a finite family of candidate patterns $\mathcal Z_\ell\subseteq\{0,1\}^n$ for every cell $\ell$. Then the worst-case loss value satisfies the certified upper bound
\begin{equation}
        \eta_{wc}(\epsilon)\le UB_{\rm loc}(\epsilon):=
        \max_{\ell=1,\ldots,C}
        \min_{z\in\mathcal Z_\ell} UB_{z,\ell}(\epsilon).
\label{eq:localized-ub-valid}
\end{equation}
\end{proposition}
\begin{proof}
If $UB_{\rm loc}(\epsilon)=\infty$, the statement is trivial. Otherwise, fix any admissible perturbation $\delta\in\mathcal U_\epsilon$. Since the sets $\mathcal U_\epsilon^\ell$ cover $\mathcal U_\epsilon$, there exists a cell $\ell$ such that $\delta\in\mathcal U_\epsilon^\ell$. For each $z\in\mathcal Z_\ell$, Proposition~\ref{prop:patterndecomp} gives
$\eta^\ast(\bar c+S\delta)\le \eta_z(\bar c+S\delta)\le UB_{z,\ell}(\epsilon)$.
Taking the minimum over $z$ yields
\[
\eta^\ast(\bar c+S\delta)\le\min_{z\in\mathcal Z_\ell} UB_{z,\ell}(\epsilon)\le UB_{\rm loc}(\epsilon).
\]
Finally, taking the maximum over $\delta\in\mathcal U_\epsilon$ gives~\eqref{eq:localized-ub-valid}.
\end{proof}

This construction is computationally simple for polyhedral cells.  Suppose
$\mathcal U_\epsilon^\ell=\operatorname{conv}\{\delta^{\ell 1},\ldots,\delta^{\ell r_\ell}\}$.
For fixed $z$, the effective domain of $\eta_z$ is polyhedral, and by the dual representation in
Lemma~\ref{lem:vzdual} the function $\eta_z(\bar c+S\delta)$ is convex; therefore its maximum over the polytope $\mathcal{U}_\epsilon^\ell$ is attained at a vertex. Hence the domain condition $\mathcal P_z(\bar c+S\delta)\ne\emptyset$ is checked by testing the vertices
$\delta^{\ell r}$ only, and, whenever it holds,
\begin{equation}
        UB_{z,\ell}(\epsilon)
        =
        \max_{r=1,\,\ldots,\,r_\ell}
        \eta_z(\bar c+S\delta^{\ell r}).
\label{eq:localized-vertex-ub}
\end{equation}
Thus each local certificate is obtained by solving the same fixed-pattern LPs as before, but only
at the vertices of a smaller cell.

The localized bound is never weaker than the global fixed-pattern bound when the partition contains the
same candidate patterns in every cell, since one may reuse the global certificate on each cell.  In
practice, it is sharper because the candidate family $\mathcal Z_\ell$ can depend on the cell and may
include regimes observed at the cell vertices or at its centroid.  An adaptive implementation starts from
$\mathcal U_\epsilon$, computes the local values in~\eqref{eq:localized-ub-valid}, bisects the cell with the
largest local upper bound, and carries the parent patterns to the children.  This produces a monotone
sequence of certified upper bounds and can be combined with the scenario lower bound exactly as in
Proposition~\ref{prop:regimegen}.
The localized certificate is especially important computationally because it avoids requiring one payment regime to be valid over the entire uncertainty set; instead, regimes are certified cell by cell over a polyhedral partition of the shock set.

\subsubsection*{Computational complexity of localized fixed-pattern certificates}\label{subsec:localized-complexity}

The localized certificate is a bounding procedure, not an exact polynomial-time algorithm for the full post-insolvency worst-case problem.  Its numerical complexity can be assessed precisely.  Suppose that the uncertainty set is covered by $C$ polyhedral cells and that cell $\ell$ has vertex set $\mathcal V_\ell$ with $q_\ell:=|\mathcal V_\ell|$.  Let $K_\ell:=|\mathcal Z_\ell|$ be the number of candidate payment patterns tested on that cell.  Once these candidate patterns have been selected, computing the localized upper bound requires at most
\[
        N_{\LP}\le \sum_{\ell=1}^{C} K_\ell q_\ell
\]
fixed-pattern LP evaluations.  Each such LP has $n$ payment variables, bound constraints $0\le p\le Z\bar p$, and at most $n$ resource constraints, so its dimension is comparable to the fixed-shock clearing LP.  Thus the certificate-generation cost is
\[
        O\!\left(N_{\LP}\,T_{\LP}(n) + N_{\rm clr}\,T_{\rm clr}(n)\right),
\]
where $T_{\LP}(n)$ denotes the cost of one fixed-pattern LP and $T_{\rm clr}(n)$ denotes the cost of one exact clearing computation used to discover candidate patterns.  In the general signed case the latter may be the MILP oracle of Theorem~\ref{thm:MILPexact}; in implementations where a monotone fixed-point clearing iteration is applicable, it is typically much cheaper.  The exact-clearing calls are used only to generate useful candidate regimes; validity of the upper bound itself follows from the fixed-pattern certificates.

For the long-only $\ell_1$ case used in the numerical section, each cell is a simplex and $q_\ell=m+1$.  After $s$ longest-edge bisections, $C=s+1$, and therefore
\[
        N_{\rm LP}=O\bigl((s+1)(m+1)K_{\max}\bigr),
        \qquad K_{\max}:=\max_\ell K_\ell,
\]
before accounting for caching.  Caching shared vertex--pattern evaluations can only reduce this number.  If candidate patterns are generated from cell vertices and centroids, the number of additional exact clearing calls grows only linearly with the number of bisections.  The cell computations are mutually independent, so the method is also directly parallelizable.  The practical trade-off is therefore explicit: increasing the number of cells tightens the certificate at the price of more small LPs and a linear number of additional clearing evaluations.

%%%%%%%%%%%%%%%%%%%%%%%%%%%%%%%%%%%%%%%%%%%%%%%%%%%%%%%%%%%%%%%%%%%%%%%%%%%%%%%
\section{Piecewise linearity and breakpoint characterization for \texorpdfstring{$\epsilon\le \epsilon_{ub}$}{epsilon <= epsilon ub}}
\label{sec:preub-pwl}

We now analyze the worst-case loss curve on the interval $[0,\epsilon_{ub}]$, where the LP-duality framework of \cite{CalafioreFracastoroProskurnikov2025ANS} is exact. For the $\ell_1$ and $\ell_\infty$ uncertainty models there are, in principle, two margins $\epsilon_{ub,1}$ and $\epsilon_{ub,\infty}$. To simplify notation we just write $\epsilon_{ub}$ for the margin corresponding to the norm currently under consideration.

This section has two purposes. First, we show that on $[0,\epsilon_{ub}]$ the worst-case loss curve is a finite upper envelope of affine functions and is therefore convex, nondecreasing, and piecewise linear. Second, we use the associated LP bases to identify the mechanisms that generate breakpoints, distinguishing changes in worst-case shock directions from changes in bank-level clearing activity.

\subsection{Worst-case loss and its dual robust representation}
\label{subsec:WorstCasel1linfty}

Fix a norm $\norm{\cdot}$ on $\R^m$ and let $\norm{\cdot}_\ast$ denote its dual norm.
For every $\epsilon \le \epsilon_{ub}$, the worst-case loss equals the optimal value of the robust dual program derived
in~\cite[Theorem~2]{CalafioreFracastoroProskurnikov2025ANS}. In this regime,
the relevant payment pattern is the all-active pattern $z=\1$, and the problem~\eqref{eq:MILP} can be replaced by the LP (Remark~\ref{rem:lp}).

Define the feasible polyhedron
\begin{equation}
X:=\bigl\{(\beta,\lambda)\in\R_+^n\times\R_+^n:\ \beta-\1+(I_n-A)\lambda\ge 0\bigr\}\subseteq\R_+^{2n}.
\label{eq:Xdef}
\end{equation}
The set $X$ is nonempty, since $(\beta,\lambda)=(\1,0)\in X$, and is pointed because it is contained in the nonnegative orthant.
In accordance with Lemma~\ref{lem:vzdual} and~\eqref{eq:wc-pattern} (see also~\cite[Theorem 2]{CalafioreFracastoroProskurnikov2025ANS}), one has
\begin{gather}
\eta^\ast(\bar c+S\delta)
=
\max_{(\beta,\lambda)\in X}
\left\{
(\1-\beta)^\top\bar p-\bar c^\top\lambda
-\delta^\top S^\top\lambda
\right\},\label{eq:eta_worst_z=1}\\
\eta_{wc}(\epsilon)
=
\max_{(\beta,\lambda)\in X}
\left\{
(\1-\beta)^\top \bar p - \bar c^\top \lambda + \epsilon \norm{S^\top\lambda}_\ast
\right\}.
\label{eq:dualwc}
\end{gather}
Here we used the fact that, for fixed $\lambda$, the robust maximization over $\norm{\delta}\le\epsilon$ contributes the support-function term
\[
\max_{\norm{\delta}\le\epsilon}(-\delta^\top S^\top\lambda)
=
\epsilon\norm{S^\top\lambda}_\ast.
\]
Any maximizer $\delta$ in this support-function problem is an adverse perturbation associated with $\lambda$; equivalently, if $\epsilon>0$, the vector $q=\delta/\epsilon$ is an adverse perturbation direction. Such a maximizer need not be unique. When $\lambda$ is the component of an optimizer of~\eqref{eq:dualwc}, this perturbation is worst-case for the robust loss problem.

 We use the signed direction sets $\mathcal Q_{\infty}$ and $\mathcal Q_{1}$ defined in~\eqref{eq:unit-balls} to express the dual norms for the $\ell_\infty$ and $\ell_1$ perturbation models.

%Let $s_j$ denote the $j$th column of $S$.
%Because $S$ may be signed, the two uncertainty norms of interest are written as finite maxima over signed shock directions. Recall that we have defined the sets
%\[
%\mathcal Q_\infty=\{-1,1\}^m,
%\qquad
%\mathcal Q_1=\{\pm e_1,\ldots,\pm e_m\}.
%\]

\subsubsection*{\texorpdfstring{$\ell_\infty$ perturbations}{ell-infinity perturbations}}
If $\norm{\delta}=\norm{\delta}_\infty$, then the dual norm is $\ell_1$ and
\[
\norm{S^\top\lambda}_\ast=\norm{S^\top\lambda}_1
=
\max_{q\in\mathcal Q_\infty} q^\top S^\top\lambda.
\]
Therefore,
\begin{equation}
\eta_{wc}(\epsilon)
=
\max_{q\in\mathcal Q_\infty}
\max_{(\beta,\lambda)\in X}
\left\{
(\1-\beta)^\top \bar p - \bar c^\top \lambda
+
\epsilon\,q^\top S^\top\lambda
\right\}.
\label{eq:dualwc-linf}
\end{equation}
Thus the worst-case loss is computed by a finite family of LPs, one for each signed joint asset-movement direction $q$. If $q$ is active in~\eqref{eq:dualwc-linf}, the corresponding adverse perturbation can be chosen as
\begin{equation}\label{eq:delta-q-relation}
\delta^\ast=-\epsilon q.
\end{equation}
If $S\ge0$, only $q=\1$ is needed, so $\delta^\ast=-\epsilon\1$.

\subsubsection*{\texorpdfstring{$\ell_1$ perturbations}{ell-1 perturbations}}
If $\norm{\delta}=\norm{\delta}_1$, then the dual norm is $\ell_\infty$ and
\[
\norm{S^\top\lambda}_\ast=\norm{S^\top\lambda}_\infty
=
\max_{q\in\mathcal Q_1} q^\top S^\top\lambda.
\]
Consequently,
\begin{equation}
\eta_{wc}(\epsilon)
=
\max_{q\in\mathcal Q_1}
\max_{(\beta,\lambda)\in X}
\left\{
(\1-\beta)^\top \bar p - \bar c^\top \lambda
+
\epsilon\,q^\top S^\top\lambda
\right\}.
\label{eq:dualwc-l1}
\end{equation}
Thus the worst-case loss is the maximum of $2m$ LP values, one for each signed single-asset direction $q\in\mathcal Q_1$. In particular, an outer shock-direction switch in the $\ell_1$ model corresponds to a change in the active signed asset direction $q=\pm e_j$, equivalently to a change in the selected asset column together with its sign. If $q$ is active in~\eqref{eq:dualwc-l1}, the corresponding adverse perturbation $\delta^\ast$ is given by the same relation~\eqref{eq:delta-q-relation} as in the $\ell_{\infty}$ case.
In the long-only case $S\ge0$, this reduces to the $m$ positive asset-column LPs~\cite{CalafioreFracastoroProskurnikov2025ANS}.

\subsection{Finite-envelope representation and candidate breakpoints}

\begin{theorem}[Piecewise-linear convexity up to $\epsilon_{ub}$]
For the $\ell_1$ and $\ell_\infty$ uncertainty models considered in this paper, the function
\[
\epsilon\mapsto \eta_{wc}(\epsilon),\qquad \epsilon\in[0,\epsilon_{ub}],
\]
is finite, continuous, nondecreasing, convex, and piecewise linear. More precisely, there exist finitely many pairs
$(\gamma_k,\mu_k)\in\R\times\R_+$, $k=1,\dots,K$, such that
\begin{equation}
\eta_{wc}(\epsilon)=\max_{k=1,\dots,K}\bigl(\gamma_k+\mu_k\epsilon\bigr),
\qquad
\epsilon\in[0,\epsilon_{ub}].
\label{eq:finite-affine-repr}
\end{equation}
The coefficients $\gamma_k$ may be chosen from the extreme points $(\beta^{(k)},\lambda^{(k)})$ of $X$ as
\begin{equation}\label{eq:gamma_k}
\gamma_k=(\1-\beta^{(k)})^\top\bar p-\bar c^\top\lambda^{(k)}.
\end{equation}
For $\ell_\infty$ perturbations one has
\begin{equation}\label{eq:mu_k_linf}
\mu_k=\norm{S^\top\lambda^{(k)}}_1,
\end{equation}
whereas for $\ell_1$ perturbations one has
\begin{equation}\label{eq:mu_k_l1}
\mu_k=\norm{S^\top\lambda^{(k)}}_\infty.
\end{equation}
\label{thm:pwl}
\end{theorem}
\begin{proof}
The polyhedron $X$ has finitely many extreme points; denote them by
\[
\operatorname{ext}(X)=\bigl\{(\beta^{(1)},\lambda^{(1)}),\dots,(\beta^{(K_X)},\lambda^{(K_X)})\bigr\}.
\]
For every $\epsilon\in[0,\epsilon_{ub}]$, the robust dual representation is exact and finite. Since the objective is linear in $(\beta,\lambda)$ for fixed $\epsilon$ after expanding the dual norm as a finite maximum over signed directions~\eqref{eq:dualwc-linf},~\eqref{eq:dualwc-l1}, an optimal solution may be chosen among the extreme points of $X$.

%Because $X$ is a nonempty pointed polyhedron, it has finitely many extreme points; denote them by
%\[
%\operatorname{ext}(X)=\bigl\{(\beta^{(1)},\lambda^{(1)}),\dots,(\beta^{(K_X)},\lambda^{(K_X)})\bigr\}.
%\]
%Since the LP value is finite and attained for every $\epsilon\in[0,\epsilon_{ub}]$, an optimum may be chosen among these extreme points.

In the $\ell_\infty$ case, applying this observation to \eqref{eq:dualwc-linf} yields
\[
\eta_{wc}(\epsilon)
=
\max_{q\in\mathcal Q_\infty}
\max_{k=1,\dots,K_X}
\left\{
(\1-\beta^{(k)})^\top\bar p-\bar c^\top\lambda^{(k)}
+
\epsilon\,q^\top S^\top\lambda^{(k)}
\right\}.
\]
Exchanging the finite maxima over $q$ and $k$ gives
\[
\eta_{wc}(\epsilon)
=
\max_{k=1,\dots,K_X}
\left\{
(\1-\beta^{(k)})^\top\bar p-\bar c^\top\lambda^{(k)}
+
\epsilon\,\norm{S^\top\lambda^{(k)}}_1
\right\}.
\]
This proves~\eqref{eq:finite-affine-repr} with the coefficients $\gamma_k$ from~\eqref{eq:gamma_k} and $\mu_k$ from~\eqref{eq:mu_k_linf}.

In the $\ell_1$ case, applying the same argument to \eqref{eq:dualwc-l1} gives
\[
\eta_{wc}(\epsilon)
=
\max_{q\in\mathcal Q_1}
\max_{k=1,\dots,K_X}
\left\{
(\1-\beta^{(k)})^\top\bar p-\bar c^\top\lambda^{(k)}
+
\epsilon\,q^\top S^\top\lambda^{(k)}
\right\}
\]
and therefore
\[
\eta_{wc}(\epsilon)=
\max_{k=1,\dots,K_X}
\left\{
(\1-\beta^{(k)})^\top\bar p-\bar c^\top\lambda^{(k)}
+
\epsilon\,\norm{S^\top\lambda^{(k)}}_\infty
\right\}.
\]
Thus \eqref{eq:finite-affine-repr} holds with $K=K_X$ and the coefficients~\eqref{eq:gamma_k} and~\eqref{eq:mu_k_l1}.

In both cases, a finite pointwise maximum of affine functions is continuous, convex, and piecewise linear. Since all slopes are nonnegative, the envelope is nondecreasing.
\end{proof}

\begingroup\revon
\begin{remark}[Dual characterization of $\epsilon_{ub}$]\label{rem:dual-eub}
The robust dual~\eqref{eq:dualwc} also detects the threshold. If $\epsilon>\epsilon_{ub}$, Proposition~\ref{prop:eub-exact} gives an admissible $\delta$ for which~\eqref{eq:feasLP} is infeasible. Since $X\neq\varnothing$, the dual objective in~\eqref{eq:eta_worst_z=1} is then unbounded above on $X$, and so is the objective of~\eqref{eq:dualwc}, which dominates it because $-\delta^\top S^\top\lambda\le\epsilon\norm{S^\top\lambda}_\ast$. For $\epsilon\le\epsilon_{ub}$, instead,~\eqref{eq:dualwc} equals $\eta_{wc}(\epsilon)\le\1^\top\bar p$. Hence $\epsilon_{ub}=\max\{\epsilon\ge0:\ \text{the value of~\eqref{eq:dualwc} is finite}\}$, and Theorem~\ref{thm:pwl} describes $\eta_{wc}$ on the whole interval on which the robust LP dual is finite.
\end{remark}
\endgroup

We call $\epsilon_0\in(0,\epsilon_{ub})$ a breakpoint (or a salient point) of $\eta_{wc}$ if the left and right derivatives are different at $\epsilon_0$. Note that such one-sided derivatives exist since the function is piecewise linear with finitely many pieces. Theorem~\ref{thm:pwl} gives all candidate locations, as formalized in the following statement.

\begin{corollary}[Candidate breakpoints and one-sided slopes]
With the notation of Theorem~\ref{thm:pwl}, every breakpoint of $\eta_{wc}$ in $[0,\epsilon_{ub})$ belongs to the finite set
\begin{equation}
\mathcal K
:=
\left\{
\frac{\gamma_\ell-\gamma_k}{\mu_k-\mu_\ell}:
 k\neq \ell,\ \mu_k\neq \mu_\ell
\right\}
\cap [0,\epsilon_{ub}).
\label{eq:breakpointset}
\end{equation}
Moreover, for every $\epsilon\in(0,\epsilon_{ub})$, the one-sided derivatives exist and are
\[
\eta'_{wc,-}(\epsilon)=\min\{\mu_k:\ \gamma_k+\mu_k\epsilon=\eta_{wc}(\epsilon)\},
\]
\[
\eta'_{wc,+}(\epsilon)=\max\{\mu_k:\ \gamma_k+\mu_k\epsilon=\eta_{wc}(\epsilon)\}.
\]
Hence $\epsilon$ is a breakpoint if and only if the active affine pieces at $\epsilon$ have at least two distinct slopes.
\label{cor:breakpointset}
\end{corollary}
\begin{proof}
A pointwise maximum of finitely many affine functions can fail to be differentiable only at intersections of active affine pieces, which gives \eqref{eq:breakpointset}. The one-sided derivative formulas are the standard formulas for convex piecewise-linear functions; see, for example, \cite[Exercise~8.31]{RockWets09}. The final statement follows immediately.
\end{proof}

\begin{remark}[From envelope enumeration to diagnostics]
Corollary~\ref{cor:breakpointset} gives a complete finite description, but explicit enumeration of all affine pieces can be prohibitive because the number of extreme points of $X$ may be large. In computations it is often preferable to generate active pieces by solving the LPs at selected radii and then to inspect the corresponding primal clearing bases. The next subsection provides the bank-level interpretation needed for this.
\end{remark}

\subsection{Breakpoint-responsible banks, activity switches, and default-set transitions}\label{sec:activity-switches}

Fix $\epsilon\in(0,\epsilon_{ub})$ and choose a worst-case perturbation $\delta^\ast(\epsilon)$ attaining $\eta_{wc}(\epsilon)$. Set
\[
c^\ast(\epsilon):=\bar c+S\delta^\ast(\epsilon),
\qquad
p^\ast(\epsilon):=p^\ast(c^\ast(\epsilon)).
\]
Since $\epsilon\le\epsilon_{ub}$, the vector $p^\ast(\epsilon)$ is the unique optimizer of the clearing LP
\begin{equation}
\max_{p\in\R^n}\ \1^\top p
\quad\text{s.t.}\quad
0\le p\le \bar p,
\qquad
(I-A^\top)p\le c^\ast(\epsilon).
\label{eq:primal-clearing-param}
\end{equation}

For bank-level diagnostics, define the sets
\begin{equation}
\begin{gathered}
\mathcal F(\epsilon):=\{i:\ p_i^\ast(\epsilon)=\bar p_i\},\quad \mathcal D(\epsilon):=\{i:\ p_i^\ast(\epsilon)<\bar p_i\}.
\\%\qquad \mathcal Z(\epsilon):=\{i:\ p_i^\ast(\epsilon)=0\},\\
\mathcal R(\epsilon):=\{i:\ p_i^\ast(\epsilon)=c_i^\ast(\epsilon)+(A^\top p^\ast(\epsilon))_i\}.
\end{gathered}\label{eq:FRZsets}
\end{equation}
Thus $\mathcal F$ records banks at full payment, $\mathcal D$ records banks at default and $\mathcal R$ records banks whose resource constraint is tight.
Since $\epsilon\le\epsilon_{ub}$, the LP-feasibility condition gives $c^\ast(\epsilon)+A^\top p^\ast(\epsilon)\geq p^\ast(\epsilon)\ge0$. Hence the positive part in~\eqref{eq:clearing} can be omitted at the maximal clearing vector, and
$p^\ast(\epsilon) = \min(\bar p,\, c^\ast(\epsilon)+A^\top p^\ast(\epsilon))$, whence
\begin{gather}
\mathcal F(\epsilon)\cup\mathcal R(\epsilon)=\{1,\ldots,n\},\label{eq:UR-cover}\\
\mathcal D(\epsilon)=\mathcal R(\epsilon)\setminus\mathcal F(\epsilon).\label{eq:default-from-UR}
\end{gather}

If $(p^\ast(\epsilon),\beta^\ast(\epsilon),\lambda^\ast(\epsilon))$ is a primal-dual optimal pair for \eqref{eq:primal-clearing-param} and its dual at $c^\ast(\epsilon)$, complementary slackness gives
\begin{equation}
\beta_i^\ast(\epsilon)\bigl(\bar p_i-p_i^\ast(\epsilon)\bigr)=0,
\qquad
\lambda_i^\ast(\epsilon)\bigl(c_i^\ast(\epsilon)+(A^\top p^\ast(\epsilon))_i-p_i^\ast(\epsilon)\bigr)=0
\label{eq:CS-bank}
\end{equation}
for every bank $i$. Under strict complementarity, positivity of $\beta_i^\ast$ and $\lambda_i^\ast$ identifies the active bank-level constraints exactly, i.e., the sets $\mathcal{F}(\epsilon)$ and $\mathcal{R}(\epsilon)$.

For a fixed $\epsilon\in(0,\epsilon_{ub})$, the active signed shock direction $q$ in~\eqref{eq:dualwc-linf} or~\eqref{eq:dualwc-l1} need not be unique. We say that an \emph{outer shock-direction switch} occurs at $\epsilon_0\in(0,\epsilon_{ub})$ if no signed direction $q\in \R^m$ can be chosen active for worst-case solutions on both sides of $\epsilon_0$, that is, if the worst-case perturbation~\eqref{eq:delta-q-relation}  cannot be chosen to depend linearly on $\epsilon$ in an open neighborhood of $\epsilon_0$.

\begin{theorem}[Breakpoint mechanisms and responsible banks]
If $\eta_{wc}$ has a breakpoint at $\epsilon_0\in(0,\epsilon_{ub})$, then at least one of the following mechanisms occurs when crossing $\epsilon_0$:
\begin{enumerate}[label=(\roman*),leftmargin=2.1em]
\item \textbf{Outer shock-direction switch:} no signed direction $q$ can be chosen active for worst-case solutions on both sides of $\epsilon_0$.
\item \textbf{Inner default-set switch:} the worst-case default set changes
$\mathcal D(\epsilon_0^-)\neq \mathcal D(\epsilon_0^+)$,
which, in view of~\eqref{eq:default-from-UR}, entail that the clearing activity switches:
\[
\bigl(\mathcal F(\epsilon_0^-),\mathcal R(\epsilon_0^-)\bigr) \neq \bigl(\mathcal F(\epsilon_0^+),\mathcal R(\epsilon_0^+)\bigr).
\]
\end{enumerate}
In particular, if the same signed shock direction $q$ is active on both sides of $\epsilon_0$, then a breakpoint must be caused by a change in the set of defaulting banks. The breakpoint-responsible banks are those whose membership in $\mathcal D$ changes when crossing $\epsilon_0$.
\label{thm:breakpointstructure}
\end{theorem}
\begin{proof}
We proceed by contradiction. We assume that, in a sufficiently small punctured neighborhood $N_0$ of $\epsilon_0$, neither mechanism occurs. Consequently, the active signed shock direction $q\in \R^m$ and the worst-case default set $\mathcal{D}$ remain constant throughout $N_0$. The contradiction is obtained below by showing that the same affine expression for $p^\ast(\epsilon)$ holds on both sides of $\epsilon_0$, which contradicts the assumption that $\epsilon_0$ is a breakpoint.
If $\mathcal D=\varnothing$ for $\epsilon\in N_0$, then every component of $p^\ast(\epsilon)=\bar p$ is constant on $N_0$.

Otherwise, if $\mathcal D\neq \varnothing$, let $k:=|\mathcal D|\leq n$ and $x(\epsilon):=(p_i^\ast(\epsilon))_{i\in\mathcal D}$ be the corresponding subvector of $p^\ast(\epsilon)$. Then
\[
\bigl(I_k-\tilde A^\top\bigr)x(\epsilon)=\rho_0+\rho_1\epsilon\quad\forall\epsilon\in N_0,
\]
where $\rho_0$ and $\rho_1$ are vectors independent of $\epsilon$, and $\tilde A=(a_{ij})_{i,j\in\mathcal D}$ is the principal submatrix of $A$ associated with $\mathcal D$. Furthermore, $0\leq x_i(\epsilon)<\bar p_i$ for all $i\in\mathcal D$ for $\epsilon\in N_0$.

We claim that $I_k-\tilde A^\top$ is nonsingular. Suppose, for the sake of contradiction, that it is singular, i.e., $\tilde A^{\top}$ (and also $\tilde A$) has eigenvalue at $1$. Since $\tilde A$ is substochastic, the Gershgorin disk theorem implies that other eigenvalues do not exceed $1$ in modulus,
and thus its spectral radius is exactly $1$. The Perron--Frobenius theorem implies the existence of a real nonnegative eigenvector $x^0\geq 0$ such that $\tilde A^\top x^0=x^0.$ For $\theta>0$ small enough, define
\[
\tilde p_i(\epsilon)=
\begin{cases}
p_i^\ast(\epsilon), & i\notin\mathcal D,\\
p_i^\ast(\epsilon)+\theta x_i^0, & i\in\mathcal D.
\end{cases}
\]
The box constraints remain satisfied for $\theta>0$ being small, because $x^0\geq 0$, $0\leq p_i^\ast(\epsilon)<\bar p_i$ for $i\in\mathcal D$. For $i\in\mathcal D$, the equality $\tilde A^\top x^0=x^0$ shows that the tight resource constraints remain satisfied. For $i\notin\mathcal D$, the payment is unchanged, while the resource term $(A^\top p)_i$ can only increase because $A\ge0$. Hence $\tilde p(\epsilon)$ is feasible for the clearing LP~\eqref{eq:primal-clearing-param}. Since $x^0\neq0$ and $x^0\ge0$, we have
\[
\1^\top \tilde p(\epsilon)>\1^\top p^\ast(\epsilon),
\]
contradicting the optimality of the maximal clearing vector $p^\ast(\epsilon)$, and thus proving that $I_k-\tilde A^{\top}$ is not singular.
Then, the subvector $x(\epsilon)$ satisfies
\[
x(\epsilon)=(I_k-\tilde A^{\top})^{-1}(\rho_0+\rho_1\epsilon) \quad\forall\epsilon\in N_0,
\]
and, consequently, the full vector $p^\ast(\epsilon)$, is affine for $\epsilon\in N_0$. It follows that
\[
\eta_{wc}(\epsilon)=\mathbf 1^\top\bar p-\mathbf 1^\top p^\ast(\epsilon)
\]
is also affine on $N_0$, contradicting the assumption that $\epsilon_0$ is a breakpoint.

Hence one of the two mechanisms must occur. If the outer shock direction does not change, then the breakpoint must be caused by a change in bank-level clearing activity. The breakpoint-responsible banks are those whose membership in $\mathcal D$ changes at $\epsilon_0$.
\color{black}
%By Corollary~\ref{cor:activityconstant} and the assumed nondegeneracy, the same clearing basis remains optimal across a neighborhood of $\epsilon_0$. Proposition~\ref{prop:piecewiseaffinep} then implies that $p^\ast(\epsilon)$ is affine through $\epsilon_0$. Therefore
%\[
%\eta_{wc}(\epsilon)=\1^\top\bar p-\1^\top p^\ast(\epsilon)
%\]
%is affine through $\epsilon_0$, contradicting the existence of a breakpoint.
%Hence at least one of the two mechanisms must occur. If the outer direction does not change, the inner activity switch is necessary, and the banks whose membership in $\mathcal F$, $\mathcal R$, or $\mathcal Z$ changes are exactly those associated with constraints entering or leaving the active basis.
\end{proof}

\begin{remark}[Actual versus candidate breakpoints]
Theorem~\ref{thm:breakpointstructure} identifies mechanisms that can generate breakpoints. Such a mechanism produces an actual breakpoint of $\eta_{wc}$ only when the one-sided slopes in Corollary~\ref{cor:breakpointset} differ. Degenerate cases may produce activity changes that leave the total-loss slope unchanged; such changes are economically informative, but they are not kinks of $\eta_{wc}$.

The default-set diagnostic gives the most transparent financial narrative: a kink occurs because, under the worst admissible shock, at least one institution crosses from full payment to default, or conversely. However, default-set changes do not exhaust all possible mechanisms: outer shock-direction switches may change the slope of the loss curve without changing $\mathcal D(\epsilon)$.
\end{remark}

In summary, for $\epsilon\le\epsilon_{ub}$, the worst-case loss curve admits the finite envelope representation
\[
\eta_{wc}(\epsilon)=\max_{k=1,\dots,K}\{\gamma_k+\mu_k\epsilon\},
\qquad \mu_k\ge0.
\]
Thus the curve is convex, nondecreasing, and piecewise linear, and every breakpoint belongs to the finite candidate set~\eqref{eq:breakpointset}. Each affine piece has a nested interpretation: the outer layer identifies an active signed shock direction $q$, the dual layer identifies an extreme point of the LP polyhedron $X$,  which through complementary slackness determines the primal worst-case default set
$\mathcal D(\epsilon)$. In the absence of an outer shock-direction switch, a breakpoint must correspond to a change in the defaulting banks (Theorem~\ref{thm:breakpointstructure}). This completes the breakpoint characterization on $[0,\epsilon_{ub}]$.

%%%%%%%%%%%%%%%%%%%%%%%%%%%%%%%%%%%%%%%%%%%%%%%%%%%%%%%%%%%%%%%%
\section{Post-insolvency geometry for \texorpdfstring{$\epsilon>\epsilon_{ub}$}{epsilon > epsilon ub}}\label{sec:beyondub}

\rev{For every $\epsilon>\epsilon_{ub}$, some admissible shock makes a bank externally insolvent (Proposition~\ref{prop:eub-exact}), the robust dual~\eqref{eq:dualwc} is $+\infty$ (Remark~\ref{rem:dual-eub}),} and the LP duality machinery of Section~\ref{sec:preub-pwl} no longer applies. The exact object then becomes the pattern decomposition developed in Section~\ref{sec:post-insolvency}:
\[
\eta^\ast(c)=\min_{z\in\{0,1\}^n}\eta_z(c),
\qquad
\eta_{wc}(\epsilon)=\rev{\sup_{\norm{\delta}\le\epsilon}}\ \min_{z\in\{0,1\}^n}\eta_z(\bar c+S\delta).
\]
This representation is explicit enough to recover a useful structural theory, while also explaining why the certified bounds of Section~\ref{sec:post-insolvency} are needed computationally.

\subsection{Patternwise geometry in the asset variable}

For each fixed pattern $z\in\{0,1\}^n$, let
\[
\mathcal D_z:=\{c\in\R^n:\ \mathcal P_z(c)\neq\varnothing\}
\]
be the effective domain of the regime-restricted loss $\eta_z$ defined in \eqref{eq:etazdef}.

\begin{proposition}[Patternwise convexity and piecewise affinity of the exact loss]
For every fixed pattern $z\in\{0,1\}^n$, the set $\mathcal D_z$ is polyhedral and the mapping
\[
c\mapsto \eta_z(c),\qquad c\in\mathcal D_z,
\]
is convex, \rev{continuous} and piecewise \rev{affine}. Consequently, the exact post-insolvency loss
\[
\eta^\ast(c)=\min_{z\in\{0,1\}^n}\eta_z(c)\rev{,\qquad \eta_z(c):=+\infty\ \text{for}\ c\notin\mathcal D_z,}
\]
is \rev{piecewise affine in the following sense: there is a finite partition of $\R^n$ into relatively open convex polyhedra (sets defined by finitely many linear equations and strict linear inequalities) on each of which $\eta^\ast$ is affine. Moreover, $\eta^\ast$ is lower semicontinuous on $\R^n$, and convex and continuous on the LP-feasible set $\mathcal F=\mathcal D_{\1}$ of Proposition~\ref{prop:eub-exact}. In general, $\eta^\ast$ is neither convex nor continuous on $\R^n$.}
%No general convexity guarantee survives for $\eta^\ast$, because it is the lower envelope of finitely many convex functions.
\label{prop:post-c-pwl}
\end{proposition}
\begin{proof}
Fix $z\in \{0,1\}^n$ and write $Z=\diag(z)$. The set of pairs $(c,p)$ satisfying $p\in\mathcal P_z(c)$ is the polyhedron
\[
\left\{(c,p)\in\R^n\times\R^n:
0\le p\le Z\bar p,
\qquad
Zp-ZA^\top p\le Zc
\right\}.
\]
Its projection onto the $c$-space is $\mathcal D_z$, hence $\mathcal D_z$ is polyhedral.

For $c\in\mathcal D_z$, Lemma~\ref{lem:vzdual} gives
\[
\eta_z(c)
=
\max_{\beta,\lambda\ge0}
\left\{
\1^\top\bar p-(Z\bar p)^\top\beta-(Zc)^\top\lambda:
\beta+(I_n-A)Z\lambda\ge\1
\right\}.
\]
The dual feasible set is a polyhedron contained in $\R_+^{2n}$ and is therefore pointed. Since the primal value is finite on $\mathcal D_z$, the dual optimum is attained at an extreme point. There are finitely many such extreme points, so $\eta_z(c)=\max_k\ell_{z,k}(c)$ is the pointwise maximum of finitely many affine functions $\ell_{z,k}$ of $c$ on $\mathcal D_z$. This proves \rev{that $\eta_z$ is convex, continuous and piecewise affine on $\mathcal D_z$}.

The identity $\eta^\ast(c)=\min_z\eta_z(c)$ follows from Proposition~\ref{prop:patterndecomp}.
Since the zero pattern $z=0$ is feasible for every $c$, the minimum is finite. \begingroup\revon
To obtain the partition, let $\mathcal H$ be the finite set of hyperplanes consisting of (a) the hyperplanes $\{a^\top c=b\}$ associated with the inequalities of a fixed H-representation of each $\mathcal D_z$, and (b) the hyperplanes $\{\ell_{z,k}(c)=\ell_{z',k'}(c)\}$ for all pairs of affine pieces whose difference is a nonconstant affine function. The faces of the hyperplane arrangement $\mathcal H$ are relatively open polyhedra partitioning $\R^n$. On each face, every inequality defining a set $\mathcal D_z$ has constant truth value, so each $\mathcal D_z$ either contains the face or is disjoint from it; and the order of all values $\ell_{z,k}(c)$ is constant. Hence on each face every $\eta_z$ is either $+\infty$ or equal to a single $\ell_{z,k}$, and so is their minimum $\eta^\ast$.

For lower semicontinuity, let $c_k\to c$ and let $q$ be a limit point of $p^\ast(c_k)$. Since $(c,p)\mapsto T_c(p)$ is continuous, $q=T_c(q)$ is a clearing vector for $c$, so $q\le p^\ast(c)$ by maximality. Therefore $\limsup_k p^\ast(c_k)\le p^\ast(c)$ componentwise and $\liminf_k\eta^\ast(c_k)\ge\eta^\ast(c)$. On $\mathcal F=\mathcal D_{\1}$, the clearing LP gives $p^\ast(c)$ (Remark~\ref{rem:lp}), so $\eta^\ast=\eta_{\1}$ there, which is convex and continuous by the first part. Convexity is not preserved by the lower-envelope operation, and continuity fails where $c$ reaches the boundary of some $\mathcal D_z$; Example~\ref{ex:two-bank-cont} below exhibits a discontinuity.
\endgroup
\end{proof}

\begingroup\revon
\begin{example}[Example~\ref{ex:two-bank}, continued: discontinuity and non-attainment]\label{ex:two-bank-cont}
Take $\epsilon=1>\epsilon_{ub}=\tfrac34$ and the admissible shocks $\delta(\xi)=(-\xi,\xi-1)^\top$, $\xi\in[0,1]$, for which $\norm{\delta(\xi)}_1=1$ and $c(\xi)=\bar c+S\delta(\xi)=(1-2\xi,\ \xi-\tfrac12)^\top$. For $\xi>\tfrac12$ the maximal clearing vector is $p^\ast(c(\xi))=(0,\ \xi-\tfrac12)^\top$: bank~1 is externally insolvent, since its available resources $1-2\xi+p_2^\ast=\tfrac12-\xi$ are negative. Hence $\eta^\ast(c(\xi))=\tfrac52-\xi\to2$ as $\xi\downarrow\tfrac12$, whereas $c(\tfrac12)=0$, $p^\ast(0)=\bar p$ and $\eta^\ast(0)=0$: the exact loss jumps down at $c=0$. Since $\eta^\ast\le\1^\top\bar p=2$, we obtain $\eta_{wc}(1)=2$. The supremum is not attained: loss $2$ requires $p^\ast=0$, hence $c\le0$, and the only admissible shock with $\bar c+S\delta\le0$ is $\delta=(-\tfrac12,-\tfrac12)^\top$, for which $c=0$ and $p^\ast=\bar p$.
\end{example}
\endgroup

Proposition~\ref{prop:post-c-pwl} pinpoints the change relative to the pre-insolvency regime. Before $\epsilon_{ub}$ the robust loss is an upper envelope of affine functions of the radius. Beyond $\epsilon_{ub}$ the statewise loss is already a lower envelope over discrete insolvency patterns, so convexity\rev{, continuity and attainment of the worst case are} no longer automatic.

\subsection{Piecewise affinity in the radius}

We now specialize again to the two uncertainty models used throughout the paper. For $\ell_1$ and $\ell_\infty$ norms, the unit ball
\[
\mathcal U:=\{\delta\in\R^m:\ \norm{\delta}\le1\}
\]
is a polytope and the radius-$\epsilon$ uncertainty set is $\epsilon\mathcal U$.

\begingroup\revon
\begin{theorem}[Post-insolvency piecewise affinity in the radius]
Assume that the uncertainty norm is either $\ell_1$ or $\ell_\infty$.
Then the exact worst-case loss curve $\eta_{wc}(\epsilon)$ is finite, nondecreasing, left-continuous on $(0,\infty)$, and piecewise affine with finitely many pieces: there are radii $0=\epsilon_0<\epsilon_1<\dots<\epsilon_N<\epsilon_{N+1}=+\infty$ such that $\eta_{wc}$ is affine on each open interval $(\epsilon_{k-1},\epsilon_k)$, $k=1,\ldots,N+1$. At a radius $\epsilon_k$ the curve may have a kink or an upward jump,
\[
\eta_{wc}(\epsilon_k)=\lim_{\epsilon\uparrow\epsilon_k}\eta_{wc}(\epsilon)\le\lim_{\epsilon\downarrow\epsilon_k}\eta_{wc}(\epsilon),
\]
and jumps can occur only at radii $\epsilon_k\ge\epsilon_{ub}$. Every affine piece has nonnegative slope, and $\eta_{wc}$ need not be convex beyond $\epsilon_{ub}$.
\label{thm:post-radius-pwl}
\end{theorem}
\begin{proof}
\emph{Monotonicity.} Let $0\le \epsilon_1\le \epsilon_2$ and denote by
$\mathcal U_\epsilon:=\{\delta\in\mathbb R^m:\ \norm{\delta}\le\epsilon\}$ the radius-$\epsilon$
uncertainty set. Since $\mathcal U_{\epsilon_1}\subseteq\mathcal U_{\epsilon_2}$,
\[
\eta_{wc}(\epsilon_1)
=
\sup_{\delta\in\mathcal U_{\epsilon_1}}\eta^\ast(\bar c+S\delta)
\le
\sup_{\delta\in\mathcal U_{\epsilon_2}}\eta^\ast(\bar c+S\delta)
=
\eta_{wc}(\epsilon_2).
\]
In particular, every affine piece of $\eta_{wc}$ has nonnegative slope.

\emph{Left continuity.} Let $g(\delta):=\eta^\ast(\bar c+S\delta)$, which is lower semicontinuous by Proposition~\ref{prop:post-c-pwl}, and let $\epsilon_0>0$. By monotonicity, $\lim_{\epsilon\uparrow\epsilon_0}\eta_{wc}(\epsilon)$ is the supremum of $g$ over the open ball $\{\norm{\delta}<\epsilon_0\}$. For $\norm{\delta}=\epsilon_0$, lower semicontinuity gives $g(\delta)\le\liminf_{t\uparrow1}g(t\delta)$, so the suprema over the open and the closed ball coincide, and $\eta_{wc}(\epsilon_0)=\lim_{\epsilon\uparrow\epsilon_0}\eta_{wc}(\epsilon)$. Together with monotonicity, this shows that $\eta_{wc}$ can only jump upward, from the right.

\emph{Piecewise affinity.} The preimages under $\delta\mapsto\bar c+S\delta$ of the cells in Proposition~\ref{prop:post-c-pwl} are relatively open polyhedra $Q_1,\dots,Q_M$ partitioning $\R^m$, and $g(\delta)=\alpha_r+b_r^\top\delta$ on $Q_r$. Therefore
\begin{equation}
\eta_{wc}(\epsilon)
=
\max_{r=1,\dots,M}\psi_r(\epsilon),
\qquad
\psi_r(\epsilon):=
\sup\{\alpha_r+b_r^\top\delta:\ \delta\in\epsilon\mathcal U\cap Q_r\},
\label{eq:post-radius-LPpieces}
\end{equation}
with $\psi_r(\epsilon)=-\infty$ if $\epsilon\mathcal U\cap Q_r=\varnothing$. Write $\mathcal U=\{\delta:\ H\delta\le h\}$ and $Q_r=\{\delta:\ E_r\delta=e_r,\ G_r\delta<f_r\}$. The set $K_r:=\{(\delta,\epsilon):\ \epsilon\ge0,\ H\delta\le\epsilon h,\ \delta\in Q_r\}$ is convex, its projection $I_r$ onto the $\epsilon$-axis is an interval (the effective domain of $\psi_r$), and $\psi_r$ is concave on $I_r$. Let $\bar K_r$ be obtained from $K_r$ by relaxing the strict inequalities to non-strict ones; if $K_r\neq\varnothing$, then $\bar K_r=\operatorname{cl}K_r$. For $\epsilon$ in the interior of $I_r$, the affine set $\R^m\times\{\epsilon\}$ meets the relative interior of $K_r$~\cite[Theorem~6.6]{Rockafellar1970}, hence its intersection with $\bar K_r$ is the closure of its intersection with $K_r$~\cite[Corollary~6.5.1]{Rockafellar1970}, and
\[
\psi_r(\epsilon)=\max\{\alpha_r+b_r^\top\delta:\ H\delta\le\epsilon h,\ E_r\delta=e_r,\ G_r\delta\le f_r\}.
\]
This is the value of an LP whose right-hand side depends affinely on the scalar parameter $\epsilon$, so it is piecewise affine with finitely many pieces~\cite{Gal1995}. Hence each $\psi_r$ is affine on each of finitely many open intervals that cover $[0,\infty)$ up to finitely many points. Adding the finitely many points at which two of these affine pieces cross, the maximum in~\eqref{eq:post-radius-LPpieces} is affine between consecutive points of the resulting finite set $\{\epsilon_1,\dots,\epsilon_N\}$.

\emph{Location of jumps, finiteness, convexity.} On $[0,\epsilon_{ub}]$ the curve is continuous by Theorem~\ref{thm:pwl}, so jumps occur only at radii $\epsilon_k\ge\epsilon_{ub}$ (a jump at $\epsilon_{ub}$ itself is possible, from the right). Finiteness follows from $0\le\eta_{wc}(\epsilon)\le\1^\top\bar p$. The possible failure of convexity follows from Proposition~\ref{prop:post-c-pwl}: $g$ is generally a lower envelope of patternwise convex functions, and maximizing a nonconvex piecewise-affine function over expanding polytopes does not preserve the convex upper-envelope structure of Theorem~\ref{thm:pwl}.
\end{proof}

\begin{example}[An upward jump]\label{ex:jump}
Let two banks owe one unit to each other, $\bar c=(0.6,0.6)^\top$, and let a single asset be held by bank~1 only, $S=e_1$, so that $\eta_{wc}(\epsilon)=\eta^\ast(\bar c-\epsilon e_1)$ for both norms. Here $\epsilon^\ast=0.6$ and, as in Example~\ref{ex:two-bank}, $\epsilon_{ub}=1.2$. The maximal clearing vector is $p^\ast=(1,1)^\top$ for $\epsilon\le0.6$, $p^\ast=(1.6-\epsilon,1)^\top$ for $0.6<\epsilon\le1.2$, and $p^\ast=(0,0.6)^\top$ for $\epsilon>1.2$: beyond $\epsilon_{ub}$ bank~1 becomes externally insolvent and the liability cycle can no longer sustain itself. The worst-case loss therefore jumps from $\eta_{wc}(1.2)=0.6$ to $\lim_{\epsilon\downarrow1.2}\eta_{wc}(\epsilon)=1.4$. A jump of this kind is a cliff effect: an arbitrarily small increase in stress beyond the threshold produces a discrete increase in systemic loss.
\end{example}
\endgroup

\begin{remark}[Fixed-pattern bounds vs. the exact min--max formulas]
For each fixed pattern $z$, the dual formula \eqref{eq:UBz-dual} gives a convex piecewise-linear upper bound $UB_z(\epsilon)$. It is tempting to combine these bounds as a min--max expression over patterns and dual extreme points. In general, however,
\[
\rev{\sup_{\delta}}\min_z \eta_z(\bar c+S\delta)
\neq
\min_z\max_{\delta}\eta_z(\bar c+S\delta),
\]
so the fixed-pattern duals are exact for upper bounds, not for the exact robust value unless a single pattern is known to be optimal throughout the relevant uncertainty set. This is why Theorem~\ref{thm:post-radius-pwl} uses a polyhedral refinement of the exact lower envelope, whereas Section~\ref{sec:post-insolvency} uses fixed patterns as certified upper approximations.
\end{remark}

\subsection*{Implications for robust computation}

\begingroup\revon
\begin{corollary}[Piecewise-affine certificates beyond \texorpdfstring{$\epsilon_{ub}$}{epsilon ub}]
Assume that the uncertainty norm is either $\ell_1$ or $\ell_\infty$. Then the certified lower and upper bounds developed in
Section~\ref{sec:post-insolvency} inherit the finite piecewise-affine structure of the post-insolvency loss geometry:
\begin{enumerate}[label=(\roman*),leftmargin=2.1em]
\item for every fixed pattern $z$, the robust domain condition of Proposition~\ref{prop:patternub} holds on an interval of radii of the form $[0,\bar\epsilon_z]$ or $[0,\infty)$ (or on no radius), on which $UB_z(\epsilon)$ is convex, nondecreasing and piecewise affine, and $UB_z(\epsilon)=+\infty$ beyond it;
\item for every finite family $\mathcal Z\subseteq\{0,1\}^n$, the finite-family upper bound
\[
UB_{\mathcal Z}(\epsilon)=\min_{z\in\mathcal Z}UB_z(\epsilon)
\]
is nondecreasing and piecewise affine with finitely many pieces, but need not be convex, and it may jump upward at the radii $\bar\epsilon_z$ where a pattern of $\mathcal Z$ loses robust feasibility;
\item if the scenario set consists of scaled directions, $\delta^{(k)}(\epsilon)=\epsilon d^{(k)}$, then the corresponding scenario lower bound
\[
LB_\Delta(\epsilon)=\max_k \eta^\ast(\bar c+S\epsilon d^{(k)})
\]
is piecewise affine with finitely many pieces in $\epsilon$, possibly with jumps;
\item the regime-generation procedure of Proposition~\ref{prop:regimegen} therefore constructs monotone certified lower and upper piecewise-affine approximations to the exact curve $\eta_{wc}(\epsilon)$.
\end{enumerate}
\label{cor:post-computation}
\end{corollary}

\begin{proof}
Part~\textit{(i)}: if the robust domain condition holds at $\epsilon_2$, it holds at every $\epsilon_1\le\epsilon_2$ because $\mathcal U_{\epsilon_1}\subseteq\mathcal U_{\epsilon_2}$, and the set of such radii is closed because $\mathcal D_z$ is closed. On it, the fixed-pattern dual representation~\eqref{eq:UBz-dual} applies: for $\ell_1$ and $\ell_\infty$ uncertainty, the support function reduces to a finite maximum over signed directions, so $UB_z$ is a finite maximum of affine functions of $\epsilon$ with nonnegative slopes.

Part~\textit{(ii)} follows because a finite minimum of nondecreasing piecewise-affine functions, each finite on an interval $[0,\bar\epsilon_z]$ and $+\infty$ beyond it, is nondecreasing and piecewise affine, although convexity and continuity are generally not preserved.

For part~\textit{(iii)}, the line $\{\bar c+S\epsilon d^{(k)}:\ \epsilon\ge0\}$ meets each cell of the partition in Proposition~\ref{prop:post-c-pwl} in a relatively open interval or a point, so each map
\[
\epsilon\mapsto \eta^\ast(\bar c+S\epsilon d^{(k)})
\]
is affine on finitely many intervals covering $[0,\infty)$ up to finitely many points. Taking the maximum over finitely many scenarios preserves this property. Part~\textit{(iv)} follows by combining parts~\textit{(ii)--(iii)} with the monotonicity and gap certificates in Proposition~\ref{prop:regimegen}.
\end{proof}
\endgroup

Corollary~\ref{cor:post-computation} explains the computational role of Section~\ref{sec:post-insolvency}. Even though convexity \rev{and continuity} may be lost beyond $\epsilon_{ub}$, the exact robust loss is nondecreasing and retains finite \rev{piecewise-affine} geometry, and the proposed lower and upper bounds are \rev{piecewise-affine} objects that can track this geometry with certified gaps.

%%%%%%%%%%%%%%%%%%%%%%%%%%%%%%%%%%%%%%%%%%%%%
\section{Numerical experiments}\label{sec:numerics}

This section illustrates the computational procedures developed above on synthetic financial networks.
The goal is to illustrate and validate three aspects of the theory: exact pointwise clearing, the pre-insolvency piecewise-linear formulas, and the post-insolvency certified bounding procedures. All numerical instances in this section have \emph{long-only} common-asset exposures, that is,
$ S\ge 0$. The more general signed-exposure case is covered by the theoretical formulations in the previous sections, but for clarity of exposition we use long-only exposures in the numerical tests. All reported losses are normalized by the total nominal interbank liabilities $\1^\top\bar p$. Unless explicitly stated otherwise, the perturbation set is the $\ell_1$ $\epsilon$-ball. \rev{The insolvency-resilience threshold $\epsilon_{ub}$ is computed exactly by Proposition~\ref{prop:eub-exact} ($m$ LPs, since $S\ge0$), and all radii in the tables and figures are reported relative to this value.} The experiments were run in Python~3.13 using \texttt{scipy.optimize.linprog} and, for MILP checks, \texttt{scipy.optimize.milp}, both with the HiGHS backend. The scripts, generated data files, tables, and figures are included in a reproducibility package available at \href{https://github.com/beppe969/MILPClearing}{\texttt{MILP\_Clearing\_code}}.

\subsection{Experimental instances and reported quantities}

Table~\ref{tab:instances} lists the five instances used in the experiments. The column ``liabilities''  indicates the type
of directed liability-network generator used in the test.  We use ER$(d)$ for an Erd\H{o}s--R\'{e}nyi directed graph with nominal edge density $d$, and CP$(d)$ for a core--periphery directed graph with the same baseline density parameter. In a CP instance the first approximately $20\%$ of nodes are declared core nodes. Core--core links are sampled with the largest probability, periphery--periphery links with the smallest probability, and core--periphery links with intermediate probabilities. After the edge set is sampled, all positive liabilities are drawn from a log-normal distribution and rescaled so that the total interbank liability is of the same order as the number of banks. The \texttt{D-post} instance is not random; it is a fixed six-bank scalar-shock example chosen to make the post-insolvency geometry visible.

The nominal external vector $\bar c$ is constructed so that the nominal system is default-free. In particular, the generator first compensates each bank for the positive part of $\bar p_i-(A^\top\bar p)_i$ and then adds a small random buffer. Thus the no-default condition $\bar c+A^\top\bar p\ge \bar p$ holds before any asset-price perturbation. The exposure matrix $S$ is then generated with nonnegative entries. In the random instances, larger exposures are assigned to banks that are relatively more vulnerable under full payment. In the scalar \texttt{D-post} instance, the nonnegative exposure vector is fixed together with the liabilities and nominal external positions. The seed column in Table~\ref{tab:instances} identifies the random seed used by the code.

\rev{Three} radii are shown for each instance. The default-resilience margin $\epsilon^\ast$ is the largest radius below which every admissible shock leaves the system default-free. The insolvency-resilience \rev{threshold $\epsilon_{ub}$ of Proposition~\ref{prop:eub-exact}} is the largest radius for which every admissible shock keeps the linear clearing formulation feasible. \rev{For comparison, the column $\underline\epsilon_{ub}$ reports the certified lower bound~\eqref{eq:epsub-lb} obtained from a single LP. The two values coincide for \texttt{D-post}, which has a single asset, and for \texttt{L-353}; on \texttt{D-val}, \texttt{D-pre} and \texttt{M-bound} the single LP underestimates the LP-solvable range by $68\%$, $10\%$ and $5.5\%$, respectively.} These numbers are not comparable across instances as absolute economic quantities, because they depend on the scaling of $S$; their role is to mark the zero-loss region and the boundary of the LP-solvable region on the plots.

\begin{table}[t]
\centering
\scriptsize
\caption{Synthetic instances used in the numerical experiments. ER$(d)$ denotes an Erd\H{o}s--R\'{e}nyi liability graph with density parameter $d$, and CP$(d)$ denotes a core--periphery liability graph with density parameter $d$. All exposure matrices satisfy $S\ge 0$. The radii are computed for the $\ell_1$ perturbation set, and for a single realization of the network with the indicated seed. \rev{The column $\underline\epsilon_{ub}$ is the single-LP lower bound~\eqref{eq:epsub-lb}; $\epsilon_{ub}$ is the exact threshold~\eqref{eq:epsub}.}}
\label{tab:instances}
{\setlength{\tabcolsep}{2.4pt}\begin{tabular}{lllllllll}
\toprule
Instance & $n$ & $m$ & liabilities & seed & $\epsilon^*$ & $\underline{\epsilon}_{ub}$ & $\epsilon_{ub}$ & role \\
\midrule
\texttt{D-val} & 8 & 3 & ER(0.35) & 11 & 0.0784 & 1.35 & 2.27 & MILP/LP/enumeration validation \\
\texttt{D-pre} & 12 & 4 & CP(0.28) & 22 & 0.0517 & 0.921 & 1.02 & pre-insolvency salient points \\
\texttt{M-bound} & 25 & 5 & CP(0.12) & 5 & 0.0992 & 0.916 & 0.967 & certified bounds/regime generation \\
\texttt{D-post} & 6 & 1 & fixed & 1 & 0.211 & 0.883 & 0.883 & post-insolvency scalar geometry \\
\texttt{L-353} & 353 & 5 & CP(0.02) & 353 & 0.0625 & 0.805 & 0.805 & hundreds-node scenario lower bounds \\
\bottomrule
\end{tabular}
}
\end{table}

A second convention is used in the tables below. The MILP internally uses a binary vector $z$: $z_i=1$ means that bank $i$ makes a positive interbank payment in the realized clearing vector, while $z_i=0$ means that bank $i$ pays zero to the interbank network. The tables report the corresponding set of zero-payment banks. The entry ``none'' means that every bank has positive interbank payment. When a table reports a number of ``regimes'', it counts distinct realized binary payment vectors $z$ observed in the corresponding experiment.

\subsection{Random pointwise clearing validation and the role of \texorpdfstring{$\epsilon_{ub}$}{epsilon ub}}

The first experiment uses \texttt{D-val} to validate pointwise clearing computations on random shocks. These shocks are not intended to approximate worst-case shocks; they are used only to test pointwise clearing computations across feasible and infeasible LP regimes. Hence, this subsection concerns \emph{shock-dependent} clearing losses,  not the robust worst-case loss curve $\eta_{wc}(\epsilon)$ in~\eqref{eq:wcdef}.

Here, for each given radius $\epsilon$, we generate 100 independent shocks on the boundary of the $\ell_1$ ball of radius $\epsilon$. For each draw we sample $g\sim\mathcal N(0,I_m)$ and set
$
        \delta=\epsilon\,{g}/{\|g\|_1}$, so that $\|\delta\|_1=\epsilon$.
For each fixed shock $\delta$, we compute the pointwise clearing loss
\[
        \eta^\ast(\bar c+S\delta)
        =
        \1^\top\bigl(\bar p-p^\ast(\bar c+S\delta)\bigr),
\]
normalized by $\1^\top\bar p$.
For every generated shock, we first attempt to compute $\eta^\ast$ by solving the ordinary fixed-shock clearing LP in
\eqref{eq:feasLP}.
%\[
%\begin{array}{ll}
%\displaystyle \min_p & \1^\top(\bar p-p)\\[1mm]
%\text{s.t.} & 0\le p\le \bar p,\\
%& \bar c+S\delta+A^\top p\ge p .
%\end{array}
%\]
If this LP is feasible, its optimizer gives the maximal clearing vector and we record
the corresponding LP loss. We then compute the same pointwise clearing loss with an
independent exact clearing oracle. On the small \texttt{D-val} instance this oracle is implemented
by exhaustive enumeration of bank-level clearing regimes.
%, not by a branch-and-bound MILP solver.
More precisely, each bank is assigned to one of three states:
zero payment, full payment, or resource-bound partial payment. For a fixed regime,
the clearing equations reduce to a linear system; the code precomputes the corresponding
affine maps and evaluates all $3^8$ regimes in vectorized form. The resulting value is
the same maximal-clearing loss that would be obtained from the MILP formulation
\eqref{eq:MILP}, but the reported time is the time of this vectorized exact-regime
oracle, not the time of a generic MILP solve. If the clearing LP is infeasible, which can
occur only outside the uniform solvency guarantee, we record no LP loss for that shock
and evaluate the clearing loss using this exact oracle only.

\begin{table}[t]
\centering
\scriptsize
\caption{Random pointwise clearing validation on \texttt{D-val}. For each value of
$\epsilon/\epsilon_{ub}$, 100 random shocks are generated on the boundary of the
$\ell_1$ ball. The column ``avg. LP loss'' averages the clearing-LP loss over
LP-feasible shocks only; the column ``avg. exact loss'' averages the exact
maximal-clearing loss over all 100 shocks. The exact loss is computed by vectorized
enumeration of all $3^8$ bank-level clearing regimes. ``Avg. insolvent banks'' is the average number of banks
with negative residual resources $c_i(\delta)+(A^\top p^\ast)_i<0$. ``LP feasible''
is the fraction of the 100 shocks for which the clearing LP is feasible.}
\label{tab:validation}
{\setlength{\tabcolsep}{2.4pt}\begin{tabular}{lllllll}
\toprule
$\epsilon/\epsilon_{ub}$ & avg. LP loss & avg. exact loss & avg. insolvent banks & LP feasible & LP s & exact s \\
\midrule
0.5 & 0.0173 & 0.0173 & 0 & 1 & 0.0016 & 0.00178 \\
1 & 0.0524 & 0.0524 & 0 & 1 & 0.00173 & 0.00185 \\
1.2 & 0.065 & 0.0662 & 0.01 & 0.99 & 0.00166 & 0.00179 \\
2 & 0.0854 & 0.181 & 0.62 & 0.72 & 0.00159 & 0.00186 \\
\bottomrule
\end{tabular}
}
\end{table}

Table~\ref{tab:validation} confirms that the clearing LP and the independent exact
regime computation give the same losses whenever the clearing LP is feasible; the
largest discrepancy in the underlying runs is at numerical precision. For
$\epsilon\le\epsilon_{ub}$, the clearing LP is feasible for all tested shocks, as guaranteed
by the definition of $\epsilon_{ub}$. For $\epsilon>\epsilon_{ub}$, the guarantee is no
longer uniform over the whole uncertainty ball: some individual shocks may still admit
a feasible clearing LP, while others may lead to external-sector insolvency and require
the exact clearing formulation. \rev{The clearing LP is infeasible for $1\%$ of the sampled shocks at $1.2\epsilon_{ub}$ and for $28\%$ at $2\epsilon_{ub}$.}
%The timing column labelled ``exact s'' therefore measures
%the specialized exact-regime oracle used for this small validation instance.

\subsection{Pre-insolvency loss curve and salient points}

\begin{figure}[h!tb]
\centering
\includegraphics[width=0.78\textwidth]{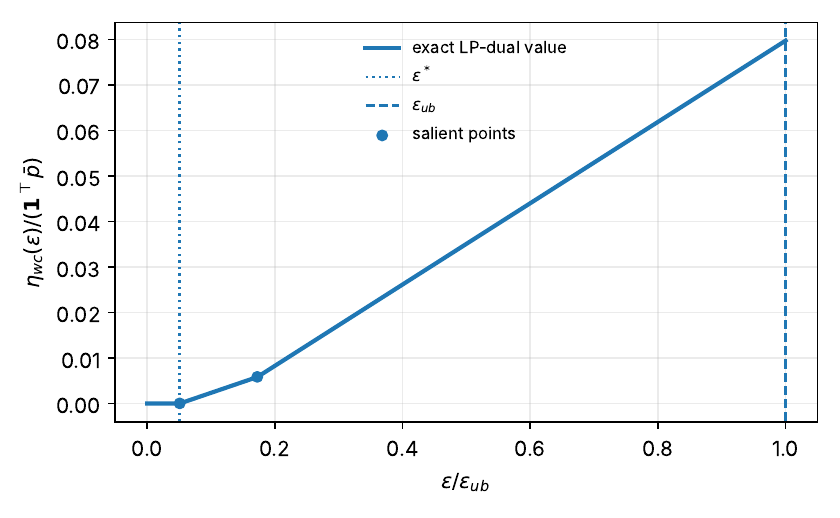}
\caption{Pre-insolvency worst-case loss curve on \texttt{D-pre}. The first salient point is $\epsilon^\ast$, where the worst-case loss becomes positive. The second salient point is caused by a change in the active adverse $\ell_1$ shock direction.}
\label{fig:pre-salient}
\end{figure}

The second experiment uses \texttt{D-pre} to inspect the exact robust worst-case loss curve on the interval $[0,\epsilon_{ub}]$, where the robust LP-dual formulation is exact. Figure~\ref{fig:pre-salient} shows the normalized worst-case loss. The curve is identically zero up to $\epsilon^\ast$ and is then convex and piecewise linear, in agreement with Theorem~\ref{thm:pwl}. The affine pieces were collected from the robust dual LPs, their upper envelope was formed, and each envelope segment was checked by resolving the robust LP at intermediate radii. \rev{The interval $[0,\epsilon_{ub}]$ is about $10\%$ longer than the interval $[0,\underline\epsilon_{ub}]$ certified by the single LP~\eqref{eq:epsub-lb} (Table~\ref{tab:instances}); no salient point lies in $(\underline\epsilon_{ub},\epsilon_{ub}]$.}

\begin{table}[h!tb]
\centering
\scriptsize
\caption{Salient-point diagnostics for \texttt{D-pre}. Both slope changes involve a switch in the outer worst-case shock direction. The activity and banks columns report the one-sided clearing-activity changes obtained by evaluating the adjacent affine segments.}
\label{tab:breakpoints}
\begin{tabular}{lllllll}
\toprule
$k$ & $\epsilon/\epsilon_{ub}$ & left slope & right slope & shock & activity & banks \\
\midrule
1 & 0.0508 & 0 & 0.565 & True & True & 4 \\
2 & 0.172 & 0.565 & 1.05 & True & True & 2,3,4,12 \\
\bottomrule
\end{tabular}

\end{table}

The two salient points are reported in Table~\ref{tab:breakpoints}. The first is the transition from the zero-loss part of the curve to the first positive-loss affine piece; the one-sided activity diagnostic identifies bank~4 as the bank whose activity changes at default onset. The second salient point is a switch in the active asset direction selected by the $\ell_1$ robust problem, and the one-sided activity sets change from the default of bank~4 to the default of banks~2, 3, and~12. Because all exposures are long-only, the adverse $\ell_1$ directions are asset-price declines. This illustrates the distinction made in Section~\ref{sec:activity-switches}: a kink in the loss curve can be driven by a change in the most damaging admissible asset-price movement and can also be accompanied by a bank-level activity switch.

\begin{figure}[htb]
\centering
\includegraphics[width=0.72\textwidth]{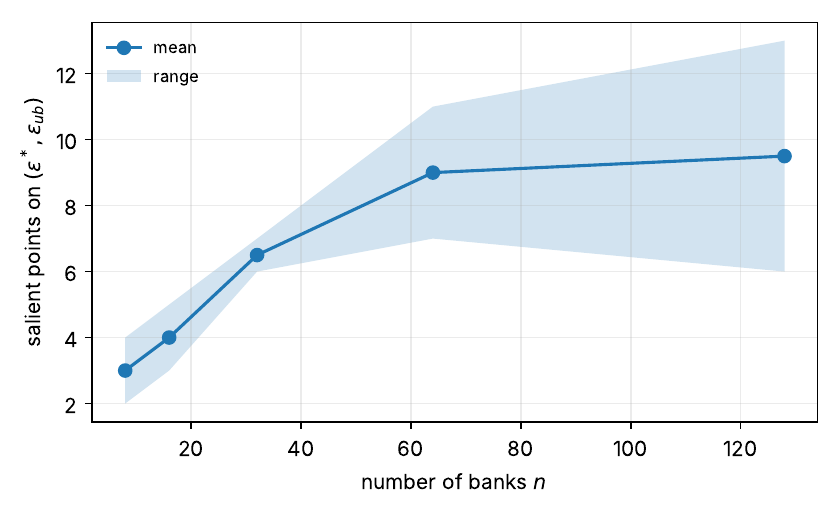}
\caption{Number of salient points in the pre-insolvency interval for the growth experiment in Table~\ref{tab:salient-growth}. The shaded band shows the range over two random seeds.}
\label{fig:salient-growth}
\end{figure}

To test how many salient points appear in typical synthetic networks, we repeated the active-line enumeration on core--periphery graphs with $n\in\{8,16,32,64,128\}$ and two random seeds for each size, keeping $m=4$. Table~\ref{tab:salient-growth} and Figure~\ref{fig:salient-growth} report the number of active affine pieces and the number of salient points on $(\epsilon^\ast,\epsilon_{ub})$. The counts increase with size in this experiment, but remain modest. This is an empirical observation about these instances, not a worst-case complexity statement.

\begin{table}[thb]
\centering
\scriptsize
\caption{Observed growth of salient points on pre-insolvency core--periphery instances. The adaptive enumeration was validated by independent LP solves; the last column gives the largest validation gap.}
\label{tab:salient-growth}
{\setlength{\tabcolsep}{3pt}\begin{tabular}{lllllll}
\toprule
$n$ & pieces mean & pieces max & salient mean & salient max & time s & max validation gap \\
\midrule
8 & 4.5 & 5 & 3 & 4 & 0.123 & 1.11e-15 \\
16 & 5.5 & 6 & 4 & 5 & 0.12 & 2.22e-15 \\
32 & 8 & 9 & 6.5 & 7 & 0.229 & 3.33e-15 \\
64 & 10.5 & 12 & 9 & 11 & 0.608 & 1.47e-14 \\
128 & 10.5 & 14 & 9.5 & 13 & 1.34 & 4.57e-14 \\
\bottomrule
\end{tabular}
}
\end{table}

%\clearpage
\subsection{Post-insolvency geometry on a scalar long-only instance}

\begin{figure}[!h]
\centering
\includegraphics[width=0.74\textwidth]{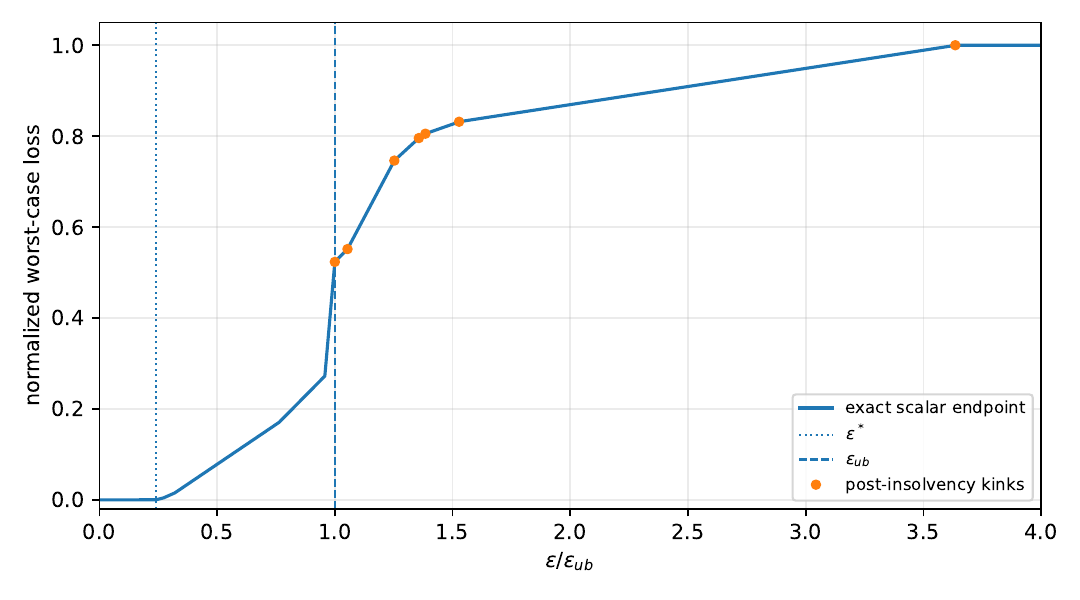}
\caption{Scalar post-insolvency example \texttt{D-post} with long-only exposures, computed by exact scalar regime enumeration at the adverse endpoint $\delta=-\epsilon$. The markers indicate the post-insolvency kinks listed in Table~\ref{tab:post-slopes}. The loss curve is nondecreasing and piecewise linear, but after $\epsilon_{ub}$ its slope can decrease, so the curve need not be convex. \rev{On this instance the curve has no jumps, which we verified by exact regime enumeration on a fine grid and at $\epsilon_k\pm10^{-6}$ around each kink.}}
\label{fig:post-scalar}
\end{figure}

\begingroup
The \texttt{D-post} instance has one asset, that is $S\in \R^n$ and $S\geq 0$. Therefore the adverse endpoint of the $\ell_1$ uncertainty interval is the price decline $\delta=-\epsilon$. The robust scalar loss curve is computed by exact one-dimensional regime enumeration along this endpoint, followed by exact clearing on each affine regime. Figure~\ref{fig:post-scalar} shows the resulting normalized loss curve and marks the post-insolvency kinks reported in Table~\ref{tab:post-slopes}. As before, the curve is zero up to $\epsilon^\ast$. Once $\epsilon$ passes $\epsilon_{ub}$, external-sector insolvency \rev{occurs at the adverse shock} and the convexity guarantee of the pre-insolvency LP formulation no longer applies. The curve remains nondecreasing, but its slope may decrease.
\endgroup

\begin{table}[!h]
\centering
\scriptsize
\caption{Post-insolvency kinks in \texttt{D-post}. Slopes are normalized consistently with Figure~\ref{fig:post-scalar}, i.e., they are slopes of normalized loss with respect to $\epsilon/\epsilon_{ub}$. The last two columns list zero-payment banks on the two sides of the kink.}
\label{tab:post-slopes}
\begin{tabular}{ccccc}
\toprule
\rev{$\epsilon/\epsilon_{ub}$} & \rev{left slope} & \rev{right slope} & \rev{left zero-pay} & \rev{right zero-pay} \\
\midrule
\rev{1} & \rev{5.91} & \rev{0.524} & \rev{none} & \rev{$\{2\}$} \\
\rev{1.054} & \rev{0.524} & \rev{0.978} & \rev{$\{2\}$} & \rev{$\{2\}$} \\
\rev{1.253} & \rev{0.978} & \rev{0.476} & \rev{$\{2\}$} & \rev{$\{2,4\}$} \\
\rev{1.357} & \rev{0.476} & \rev{0.346} & \rev{$\{2,4\}$} & \rev{$\{2,4,5\}$} \\
\rev{1.385} & \rev{0.346} & \rev{0.184} & \rev{$\{2,4,5\}$} & \rev{$\{2,4,5,6\}$} \\
\rev{1.528} & \rev{0.184} & \rev{0.0798} & \rev{$\{2,4,5,6\}$} & \rev{$\{1,2,4,5,6\}$} \\
\rev{3.636} & \rev{0.0798} & \rev{0} & \rev{$\{1,2,4,5,6\}$} & \rev{$\{1,2,3,4,5,6\}$} \\
\bottomrule
\end{tabular}

\end{table}

\begingroup
Table~\ref{tab:post-slopes} reports the kink locations and slopes in the same normalized coordinates as Figure~\ref{fig:post-scalar}. The visually salient change near $1.5\epsilon_{ub}$ corresponds to the exact kink at $\epsilon/\epsilon_{ub}=1.528$, where the zero-payment set expands from $\{2,4,5,6\}$ to $\{1,2,4,5,6\}$. Some kinks change the zero-payment set, while the kink at $1.054\epsilon_{ub}$ changes the active affine clearing regime without changing the zero-payment bank set. These effects are consistent with Theorem~\ref{thm:post-radius-pwl}: beyond $\epsilon_{ub}$, the exact loss is governed by a finite family of payment regimes and need not be convex.
\endgroup

\subsection{Certified bounds and regime generation}

The \texttt{M-bound} instance is used to illustrate the certified lower and upper bounds developed in Section~\ref{sec:post-insolvency}.
Since all numerical instances have long-only exposures $S\ge 0$, the maximal clearing vector is monotone in the external-resource vector $c$ and the loss is nonincreasing in $c$. Hence the worst long-only $\ell_1$ shocks may be restricted, without loss, to price declines. We therefore sample shocks of the form
\[
        \delta=-\epsilon w,\qquad w\in \R^m,\;\;\;w\ge 0,\qquad \mathbf 1^\top w=1 .
\]
For each displayed radius, the scenario set contains all single-asset vertices $-\epsilon e_j$, pairwise edge points, and 1500 Dirichlet boundary samples with several concentration parameters. The zero shock is also included for regime identification. Each scenario is evaluated with the exact clearing oracle, giving the deterministic lower bound
\[
        LB_\Delta(\epsilon)
        =
        \max_{\delta\in\Delta}\eta^\ast(\bar c+S\delta).
\]
The upper bound is computed from the fixed-pattern certificates of Proposition~\ref{prop:patternub}. For a fixed payment regime $z$, we first check the robust-domain condition required for $UB_z(\epsilon)$ to be finite. In the present long-only $\ell_1$ setting, this condition is checked at $0$ and at the negative vertices $-\epsilon e_j$, $j=1,\ldots,m$. This certifies the whole $\ell_1$ ball: positive price moves only increase $\bar c+S\delta$, and fixed-pattern feasibility is monotone in $c$, while convexity of the effective domain covers convex combinations of the tested vertices. This implements the robust-domain requirement in Proposition~\ref{prop:patternub} without using the overly conservative componentwise test that would require all banks to suffer their individual worst deterioration simultaneously \rev{(the same distinction as between~\eqref{eq:epsub} and~\eqref{eq:epsub-lb})}. Whenever the domain condition is satisfied, $UB_z(\epsilon)$ is computed by the fixed-pattern robust dual problem. Since $S\ge 0$, this reduces to $m$ LPs, one for each adverse single-asset direction.

The candidate family $\mathcal Z$ is enlarged in two ways. First, we add the payment regimes realized by the sampled scenarios. Second, \rev{at every radius beyond $\epsilon_{ub}$} we generate additional robust candidate regimes by a greedy fixed-regime search: starting from the all-zero-payment regime, a bank is activated only if the resulting regime still satisfies the robust-domain condition and improves the value of $UB_z(\epsilon)$. This greedy step does not define a new upper bound; it only supplies more candidate regimes to the same bound
\[
        UB_{\mathcal Z}(\epsilon)=\min_{z\in\mathcal Z}UB_z(\epsilon)
\]
from Section~\ref{subsec:gap-certificate}.

\begin{figure}[t]
\centering
\includegraphics[width=0.95\textwidth]{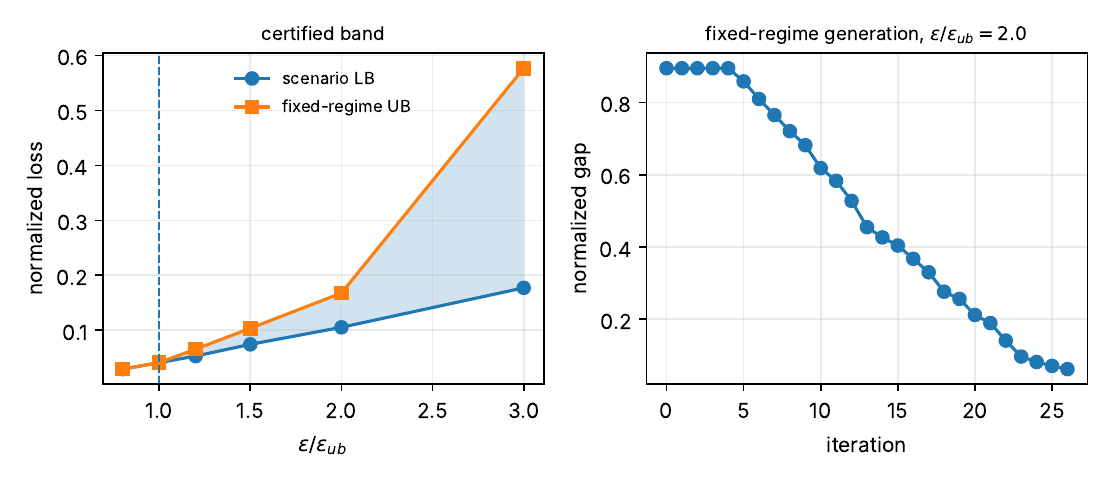}
\caption{Certified lower and upper bounds on \texttt{M-bound} using the fixed-pattern method of Proposition~\ref{prop:patternub}. Left: scenario lower bound and fixed-regime upper bound. Right: monotone reduction of the certified gap at $\epsilon/\epsilon_{ub}=2$ as additional fixed regimes are added to $\mathcal Z$.}
\label{fig:bounds}
\end{figure}

\begin{table}[t]
\centering
\scriptsize
\caption{Certified bounds on \texttt{M-bound}. Losses are normalized by $\mathbf 1^\top\bar p$. The lower bound is the best exact clearing loss over the sampled scenarios. The upper bound is $UB_{\mathcal Z}(\epsilon)=\min_{z\in\mathcal Z}UB_z(\epsilon)$, where every finite $UB_z$ is computed by the fixed-pattern robust dual problem. The column ``finite regimes'' counts candidate regimes satisfying the robust-domain condition over the whole adverse $\ell_1$ simplex. The relative gap is $(UB-LB)/UB$ when $UB>0$.
 1536 scenarios are used.}
\label{tab:bounds}
{\setlength{\tabcolsep}{3pt}\begin{tabular}{lllllll}
\toprule
$\epsilon/\epsilon_{ub}$ & LB & UB & rel. gap & candidate regimes & finite regimes & realized regimes \\
\midrule
0.8 & 0.0294 & 0.0294 & 0 & 2 & 2 & 1 \\
1 & 0.0412 & 0.0412 & 0 & 3 & 3 & 2 \\
1.2 & 0.0531 & 0.0656 & 0.191 & 26 & 25 & 2 \\
1.5 & 0.0744 & 0.104 & 0.282 & 27 & 24 & 3 \\
2 & 0.105 & 0.168 & 0.371 & 27 & 23 & 4 \\
3 & 0.177 & 0.577 & 0.693 & 21 & 11 & 10 \\
\bottomrule
\end{tabular}
}
\end{table}

\begin{table}[t]
\centering
\scriptsize
\caption{Fixed-regime generation trace for \texttt{M-bound} at $\epsilon/\epsilon_{ub}=2$. The lower bound is fixed by the sampled scenario set, while the upper bound decreases monotonically as additional candidate regimes are added to $\mathcal Z$. 1536 scenarios are used.}
\label{tab:regime-trace}
\begin{tabular}{llllll}
\toprule
iter. & LB & UB & gap & candidate regimes & finite regimes \\
\midrule
0 & 0.105 & 1 & 0.895 & 1 & 1 \\
1 & 0.105 & 1 & 0.895 & 2 & 1 \\
2 & 0.105 & 1 & 0.895 & 3 & 1 \\
3 & 0.105 & 1 & 0.895 & 4 & 1 \\
4 & 0.105 & 1 & 0.895 & 5 & 1 \\
5 & 0.105 & 0.963 & 0.858 & 6 & 2 \\
6 & 0.105 & 0.915 & 0.809 & 7 & 3 \\
7 & 0.105 & 0.87 & 0.765 & 8 & 4 \\
8 & 0.105 & 0.826 & 0.72 & 9 & 5 \\
9 & 0.105 & 0.787 & 0.682 & 10 & 6 \\
10 & 0.105 & 0.724 & 0.618 & 11 & 7 \\
11 & 0.105 & 0.689 & 0.583 & 12 & 8 \\
12 & 0.105 & 0.633 & 0.528 & 13 & 9 \\
13 & 0.105 & 0.56 & 0.455 & 14 & 10 \\
14 & 0.105 & 0.532 & 0.427 & 15 & 11 \\
15 & 0.105 & 0.51 & 0.404 & 16 & 12 \\
16 & 0.105 & 0.473 & 0.367 & 17 & 13 \\
17 & 0.105 & 0.436 & 0.33 & 18 & 14 \\
18 & 0.105 & 0.382 & 0.276 & 19 & 15 \\
19 & 0.105 & 0.362 & 0.257 & 20 & 16 \\
20 & 0.105 & 0.317 & 0.212 & 21 & 17 \\
21 & 0.105 & 0.295 & 0.19 & 22 & 18 \\
22 & 0.105 & 0.247 & 0.141 & 23 & 19 \\
23 & 0.105 & 0.202 & 0.097 & 24 & 20 \\
24 & 0.105 & 0.187 & 0.082 & 25 & 21 \\
25 & 0.105 & 0.177 & 0.0711 & 26 & 22 \\
26 & 0.105 & 0.168 & 0.0623 & 27 & 23 \\
\bottomrule
\end{tabular}

\end{table}

Figure~\ref{fig:bounds} and Table~\ref{tab:bounds} show that the certified bounds coincide at $0.8\epsilon_{ub}$ and at $\epsilon_{ub}$. \rev{Beyond the threshold the global fixed-regime certificate is conservative, with relative gaps of $19\%$ at $1.2\epsilon_{ub}$, $28\%$ at $1.5\epsilon_{ub}$ and $37\%$ at $2\epsilon_{ub}$}, but the enlarged candidate family still gives a substantially better certificate than the trivial all-zero-payment bound. In particular, at $2\epsilon_{ub}$ the upper bound is reduced to \rev{$0.168$}, while the lower bound is \rev{$0.105$}. Table~\ref{tab:regime-trace} shows the monotone tightening at this radius: the upper bound starts from the trivial value $1$ and decreases as more finite fixed-regime certificates are added. At $3\epsilon_{ub}$ the bound remains wider \rev{($0.577$ against $0.177$)}, reflecting the fact that few nontrivial regimes are valid uniformly over such a large uncertainty set. This is a limitation of the global fixed-pattern certificate, not of the pointwise clearing computation.

We also implemented the localized fixed-pattern certificate described in
Section~\ref{subsec:localized-pattern-ub}.  In the present long-only $\ell_1$ setting the adverse set is
\[
        \mathcal U_\epsilon^-
        =
        \{\delta=-\epsilon x:\ x\ge 0,\; \mathbf 1^\top x\le 1\}
        =
        \operatorname{conv}\{0,-\epsilon e_1,\ldots,-\epsilon e_m\}.
\]
We adaptively subdivide this simplex by longest-edge bisection.  On each cell we form a local candidate
family from the regimes realized at the cell vertices and centroid, together with the patterns inherited
from the parent cell \rev{and the componentwise intersection of the vertex and centroid regimes, as in the 353-bank experiment below. The intersection is needed on cells that straddle a regime switch, where each realized regime is infeasible at some vertex}.  A candidate regime is accepted on a cell only if the corresponding restricted LP is
feasible at all cell vertices.  The cell upper bound is then the largest fixed-pattern loss over those
vertices, and the global localized certificate is the maximum of the cell bounds.  The final reported
upper bound is the better of the global certificate and the localized one.

\begin{figure}[t]
\centering
\includegraphics[width=0.78\textwidth]{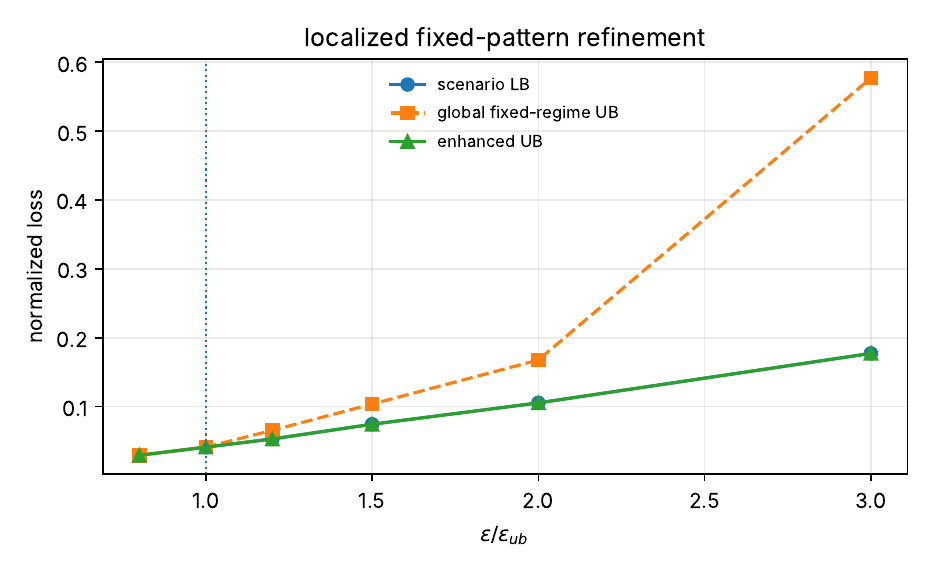}
\caption{Localized fixed-pattern refinement on \texttt{M-bound}.  The dashed curve is the global
fixed-regime upper bound, while the solid enhanced upper bound combines the global and localized
certificates.  The localized certificate closes the gap for the displayed post-insolvency radii.}
\label{fig:localized-bounds}
\end{figure}

\begin{figure}[t]
\centering
\includegraphics[width=0.78\textwidth]{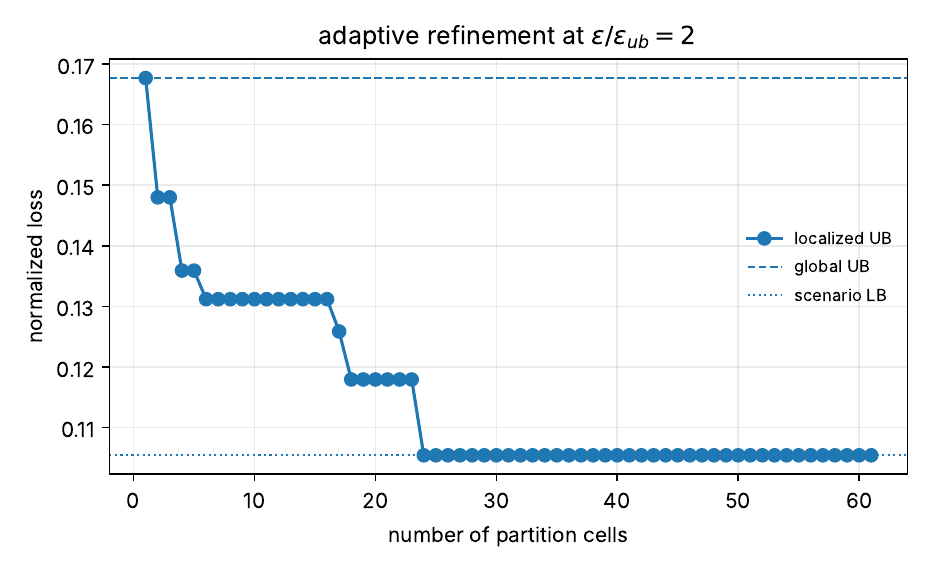}
\caption{Adaptive partition trace for \texttt{M-bound} at $\epsilon/\epsilon_{ub}=2$.  The upper bound starts from
the one-cell certificate, \rev{which already equals the global fixed-regime certificate}, drops below it after a few bisections,
and then reaches the scenario lower bound up to numerical tolerance.}
\label{fig:localized-trace}
\end{figure}

\begin{table}[t]
\centering
\scriptsize
\caption{Effect of the localized fixed-pattern refinement on \texttt{M-bound}.  Losses are normalized by
$\mathbf 1^\top\bar p$.  The column ``global UB'' is the global fixed-regime certificate from
Table~\ref{tab:bounds}.  The column ``localized UB'' is the adaptive cell-wise certificate; the enhanced
upper bound is the minimum of the two.}
\label{tab:localized-bounds}
\begin{tabular}{lllllll}
\toprule
$\epsilon/\epsilon_{ub}$ & LB & global UB & localized UB & enhanced UB & rel. gap & cells \\
\midrule
0.8 & 0.0294 & 0.0294 & -- & 0.0294 & 0 & -- \\
1 & 0.0412 & 0.0412 & -- & 0.0412 & 0 & -- \\
1.2 & 0.0531 & 0.0656 & 0.0531 & 0.0531 & 0 & 21 \\
1.5 & 0.0744 & 0.104 & 0.0744 & 0.0744 & 0 & 31 \\
2 & 0.105 & 0.168 & 0.105 & 0.105 & 2.58e-10 & 61 \\
3 & 0.177 & 0.577 & 0.177 & 0.177 & 2.27e-10 & 81 \\
\bottomrule
\end{tabular}

\end{table}

Table~\ref{tab:localized-bounds} and Figures~\ref{fig:localized-bounds}--\ref{fig:localized-trace}
show that localization removes the main conservatism of the global fixed-pattern certificate on this
instance. \rev{The localized certificate closes the gap at all four post-insolvency radii.} For example, at $\epsilon=2\epsilon_{ub}$ the global upper bound is \rev{$0.168$}, whereas the
localized certificate gives \rev{$0.105$}, matching the scenario lower bound up to numerical solver tolerances.
Therefore the sampled scenario is certified to be worst-case at this radius. At $\epsilon=3\epsilon_{ub}$,
the global upper bound \rev{$0.577$} is reduced to \rev{$0.177$}, again matching the best sampled scenario up to
solver tolerances. In both cases, the scenario value provides a certified lower bound and the localized
fixed-pattern construction provides the matching upper bound; equality of the two bounds certifies
optimality for this generated instance.

We finally report the computational effort required by the localized certificate.  The timings in Table~\ref{tab:localized-complexity} were measured in the same Python/HiGHS implementation used for the other experiments.
They are hardware-dependent, but the counters are reproducible.  \rev{The scenario lower bound uses 1536 exact clearing evaluations at each radius.}  The column ``local clearings'' counts the additional exact clearings used only to identify local candidate regimes at cell vertices and centroids.  The column ``fixed-pattern LPs'' counts the LP evaluations used to certify the localized upper bound.

\begin{table}[t]
\centering
\scriptsize
\caption{Computational effort of the localized fixed-pattern certificate on \texttt{M-bound}.  Losses are normalized by $\mathbf 1^\top\bar p$.  The localized cost is the extra cost of the cell-wise upper-bound refinement, excluding the scenario lower-bound evaluation.  1536 scenarios are used.}
\label{tab:localized-complexity}
{\setlength{\tabcolsep}{3pt}\begin{tabular}{llllllll}
\toprule
$\epsilon/\epsilon_{ub}$ & cells & local clearings & fixed-pattern LPs & scenario s & local s & total s & enhanced UB \\
\midrule
1.2 & 21 & 58 & 51 & 0.0806 & 0.0896 & 0.17 & 0.0531 \\
1.5 & 31 & 82 & 105 & 0.0877 & 0.187 & 0.275 & 0.0744 \\
2 & 61 & 154 & 236 & 0.095 & 0.411 & 0.506 & 0.105 \\
3 & 81 & 202 & 708 & 0.109 & 1.3 & 1.4 & 0.177 \\
\bottomrule
\end{tabular}
}
\end{table}

The additional computational burden is modest for this benchmark.  At $\epsilon=2\epsilon_{ub}$ the enhanced certificate uses $61$ final cells, $154$ local exact clearings, and \rev{$236$} fixed-pattern LPs; the localized upper-bound computation takes about \rev{$0.41$} seconds and closes the global-bound gap.  At $\epsilon=3\epsilon_{ub}$ the computation uses $81$ cells, \rev{$202$} local exact clearings, and \rev{$708$} LPs; the localized refinement takes about \rev{$1.3$} seconds and reduces the global upper bound from \rev{$0.577$ to $0.177$}, matching the scenario lower bound up to numerical tolerance.  These numbers illustrate the expected behavior described in Section~\ref{subsec:localized-complexity}: the cost grows with the number of cells and tested patterns, while each individual certificate remains a small LP of the same scale as a clearing LP.

\begin{figure}[t]
\centering
\includegraphics[width=0.72\textwidth]{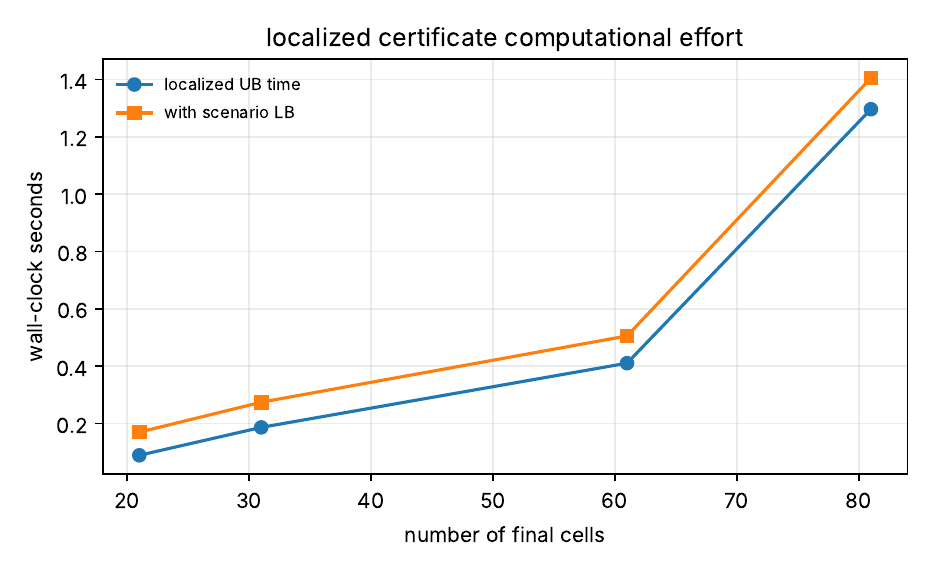}
\caption{Wall-clock effort for the localized fixed-pattern certificate on \texttt{M-bound}.  The lower curve reports only the cell-wise upper-bound refinement; the upper curve also includes the scenario lower-bound computation.}
\label{fig:localized-complexity-effort}
\end{figure}

\subsection{Scenario lower bounds on a 353-bank network}

The last experiment applies the scenario lower-bound method to \texttt{L-353}, a core--periphery network with 353 banks and five assets. At each radius, the scenario set contains all single-asset price-decline directions, random boundary scenarios, the zero shock, and a heuristic direction obtained from the pre-insolvency dual problem.
\begingroup
Each candidate shock is then re-evaluated by the monotone fixed-point iteration
\[
        p^{(0)}=\bar p,\qquad
        p^{(k+1)}=T_c\bigl(p^{(k)}\bigr),\qquad
        T_c(p):=\pos{\min(\bar p,\,c+A^\top p)},
        \quad c=\bar c+S\delta .
\]
Because $T_c$ is monotone and $T_c(\bar p)\le \bar p$, the sequence decreases componentwise and converges to the greatest fixed point of $T_c$, i.e., the maximal clearing vector~\cite{ElsingerLeharSummer2006,CalafioreFracastoroProskurnikov2025ANS}. The reported residual is the fixed-point residual $\|p-T_c(p)\|_\infty$ at termination.
\endgroup
The residuals reported in Table~\ref{tab:large-scenario} are below $3\cdot 10^{-9}$.
\begin{figure}[t]
\centering
\includegraphics[width=0.78\textwidth]{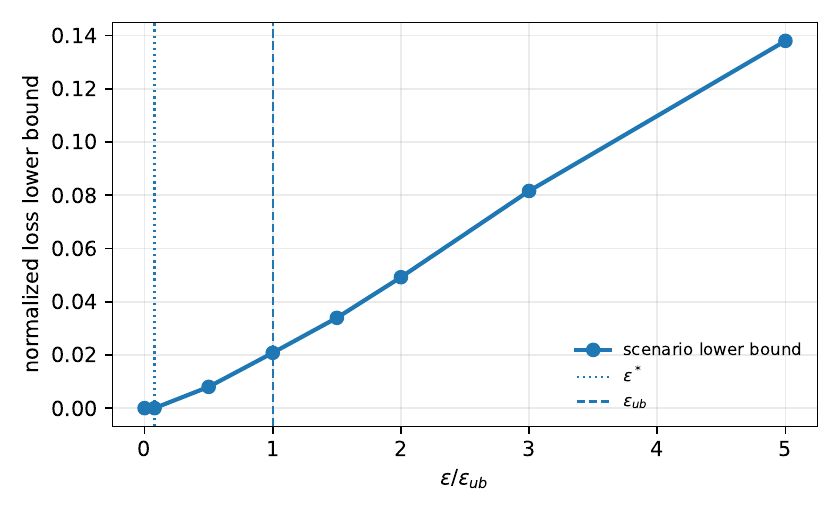}
\caption{Scenario lower bounds beyond $\epsilon_{ub}$ on the 353-bank long-only instance. The curve is a deterministic lower bound on the worst-case loss, not an exact worst-case value.}
\label{fig:large-scenario}
\end{figure}

\begin{table}[t]
\centering
\scriptsize
\caption{Scenario lower bounds on \texttt{L-353}. The column ``LP feasible at best'' reports whether the ordinary clearing LP is feasible for the best scenario found at that radius. This is a fixed-shock test and does not contradict the definition of $\epsilon_{ub}$, which is a uniform guarantee over all shocks in the ball.}
\label{tab:large-scenario}
{\setlength{\tabcolsep}{2pt}\begin{tabular}{lllllll}
\toprule
$\epsilon/\epsilon_{ub}$ & LB & scenarios & patterns & LP feasible at best & median s & res. \\
\midrule
0 & 0 & 41 & 1 & True & 0.00134 & 0 \\
0.0776 & 0 & 42 & 1 & True & 0.00117 & 0 \\
0.5 & 0.00799 & 42 & 1 & True & 0.00123 & 1.21e-09 \\
1 & 0.0208 & 42 & 2 & True & 0.00123 & 1.18e-09 \\
1.5 & 0.0339 & 42 & 3 & False & 0.00125 & 7.22e-10 \\
2 & 0.0492 & 42 & 7 & False & 0.00141 & 1.15e-09 \\
3 & 0.0816 & 42 & 10 & False & 0.00138 & 1.48e-09 \\
5 & 0.138 & 42 & 23 & False & 0.00138 & 2.17e-09 \\
\bottomrule
\end{tabular}
}
\end{table}

Figure~\ref{fig:large-scenario} shows that the scenario lower bound continues to grow beyond $\epsilon_{ub}$. Table~\ref{tab:large-scenario} also reports the number of distinct realized payment regimes among the tested scenarios. This number is one below $\epsilon_{ub}$ and grows after the insolvency radius, indicating that the post-insolvency regime structure becomes richer as the uncertainty ball expands.

We next complement the scenario lower bounds with localized fixed-pattern upper bounds.  The computation uses the adverse long-only $\ell_1$ simplex
\[
        \mathcal U_\epsilon^-=
        \operatorname{conv}\{0,-\epsilon e_1,\ldots,-\epsilon e_m\}.
\]
For the post-insolvency radii $\epsilon/\epsilon_{ub}\in\{1.5,2,3\}$ we adaptively partition this simplex by longest-edge bisection and apply the localized certificate of Section~\ref{subsec:localized-pattern-ub} on each cell.  Candidate regimes on a cell are obtained from exact clearings at the cell vertices and centroid, augmented with intersections of these local regimes.  The intersections are useful in the large network because a regime observed at one vertex need not be feasible at another vertex, while the intersection often gives a conservative pattern that remains feasible throughout the cell and is still much sharper than the all-zero pattern.  Each candidate regime is then certified by fixed-pattern LPs at the cell vertices.  The largest radius $5\epsilon_{ub}$ is kept as a scenario-based scalability stress test; producing a tight global upper certificate there would require a much finer partition and is not needed for demonstrating the upper-bound refinement.

\begin{figure}[t]
\centering
\includegraphics[width=0.74\textwidth]{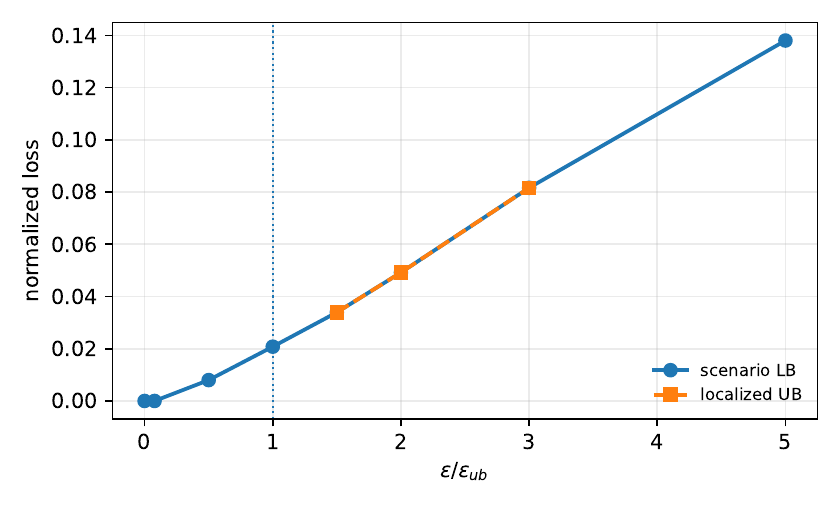}
\caption{Large-network scenario lower bounds and localized fixed-pattern upper bounds on the 353-bank instance.  The localized certificates are computed at the post-insolvency radii $1.5\epsilon_{ub}$, $2\epsilon_{ub}$, and $3\epsilon_{ub}$.}
\label{fig:large-localized-bounds}
\end{figure}

\begin{table}[t]
\centering
\scriptsize
\caption{Localized fixed-pattern certificates on the 353-bank instance.  Losses are normalized by $\mathbf 1^\top\bar p$.  The column ``local clearings'' counts exact clearing evaluations used to generate local candidate regimes; ``pattern LPs'' counts the fixed-pattern LP evaluations used to certify the upper bound.}
\label{tab:large-localized-bounds}
{\setlength{\tabcolsep}{3pt}\begin{tabular}{llllllll}
\toprule
$\epsilon/\epsilon_{ub}$ &     LB & localized UB & rel. gap &  cells &  local clearings &  pattern LPs & local s \\
\midrule
                     1.5 & 0.0339 &       0.0339 &  2.2e-10 &      5 &               19 &           50 &    1.52 \\
                       2 & 0.0492 &       0.0492 & 1.51e-10 &     35 &               92 &          350 &     9.9 \\
                       3 & 0.0816 &       0.0816 & 4.32e-10 &     66 &              172 &          789 &    21.2 \\
\bottomrule
\end{tabular}
}
\end{table}

\begin{figure}[t]
\centering
\includegraphics[width=0.74\textwidth]{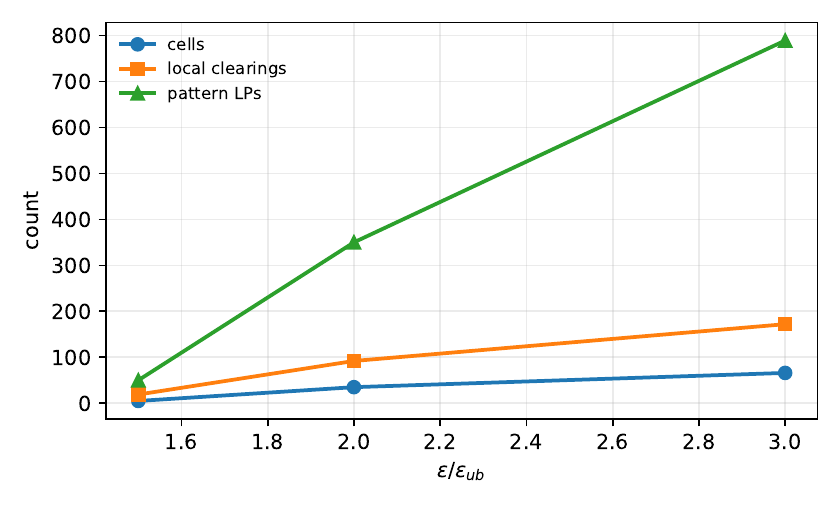}
\caption{Work counters for the localized fixed-pattern upper-bound computation on the 353-bank instance.  Each fixed-pattern LP has the scale of one clearing LP, while the exact clearings are used only to discover candidate local regimes.}
\label{fig:large-localized-effort}
\end{figure}

Table~\ref{tab:large-localized-bounds} shows that the localized upper bound matches the scenario lower bound, up to numerical tolerance, at all three certified post-insolvency radii.  Thus, for this generated 353-bank instance, the best sampled scenarios are not merely good stress scenarios: they are certified worst-case scenarios at the reported radii.  The computational effort remains moderate.  At $3\epsilon_{ub}$, the certificate uses $66$ final cells, $172$ local exact clearings, and $789$ fixed-pattern LPs.  Figure~\ref{fig:large-localized-effort} reports the corresponding work counters.

%%%%%%%%%%%%%%%%%%%%%%%%%%%%%%%%%%%%%%%%%
\section{Conclusions}\label{sec:concl}

This paper develops a certifiable computational framework for robust worst-case contagion analysis in financial networks with interbank liabilities, common asset exposures, and signed external net positions. The main difficulty is the post-insolvency regime. Before the insolvency-resilience threshold $\epsilon_{ub}$, clearing can be handled by linear programming and worst-case losses can be computed by LP duality. \rev{This threshold is computed exactly by one LP per vertex of the uncertainty ball, while a single LP based on the componentwise worst deterioration of external resources gives a certified lower bound that can be markedly conservative.} Beyond $\epsilon_{ub}$, \rev{some} admissible shocks render some banks insolvent with respect to the external sector, the LP clearing formulation \rev{becomes} infeasible \rev{for them}, and the robust stress-test problem acquires a discrete regime structure.

The first contribution is an exact MILP characterization of the maximal clearing vector for arbitrary signed external net positions. This formulation provides a clearing oracle that remains valid after external-sector insolvency appears. The second contribution is a pattern-based representation of the post-insolvency loss and a family of certified bounds: deterministic scenario lower bounds, fixed-pattern upper bounds, regime generation, and localized fixed-pattern certificates. Localization is the key tool for sharpening the upper bound: rather than requiring a single payment regime to remain valid over the entire uncertainty set, it certifies regimes cell by cell over a polyhedral partition.
The third contribution is the parametric analysis of the loss curve. For $\ell_1$ and $\ell_\infty$ price-perturbation sets, the pre-insolvency worst-case loss curve is finite, nondecreasing, convex, and piecewise linear. Its breakpoints can be traced to switches in adverse shock directions and bank-level default patterns. Beyond $\epsilon_{ub}$, monotonicity and \rev{a finite piecewise-affine structure} persist for polyhedral uncertainty sets, but convexity may fail because the exact loss becomes a lower envelope over discrete insolvency regimes. \rev{Because the exact loss is only lower semicontinuous, the curve may also jump upward, a cliff effect in which a small increase in stress produces a discrete increase in loss, and the worst case need not be attained, so certificates are in general approximate rather than exact.} This distinction clarifies why post-insolvency stress testing requires certificates rather than a single LP-duality formula.

The numerical experiments support these conclusions on synthetic benchmarks. The small and medium instances validate the clearing oracle, illustrate the pre-insolvency kink structure, and show how localized certificates sharpen global fixed-pattern upper bounds. The 353-bank core--periphery experiment demonstrates that the approach scales beyond toy networks: scenario lower bounds can be evaluated rapidly by exact clearing iteration, and localized fixed-pattern certificates provide matching upper bounds at representative post-insolvency radii.

The exact robust problem beyond $\epsilon_{ub}$ is intrinsically nonconvex, 
so certificate efficiency depends on the quality of the generated regimes and partition. 
Further work may combine the present methods with stronger branch-and-bound schemes, adaptive partitioning 
rules, or decomposition methods for large networks. It would also be natural to extend the framework 
to clearing mechanisms with bankruptcy costs, fire sales, multiple maturities, central counterparties, 
or illiquid assets. The results here provide the MILP, LP, and certificate infrastructure needed for such extensions.

\backmatter

% \bmhead{Funding}
% This work was funded by...

\bibliography{refs_revised}

\end{document}